\documentclass[
superscriptaddress,
 amsmath,amssymb,aps, prl, twocolumn]{revtex4-2}

\usepackage{graphicx}
\usepackage{dcolumn}
\usepackage{bm}
\usepackage{hyperref}

\begin{document}

\title{A mesoscopic theory for stochastic coupled oscillators}

\author{Victor Buend\'ia}
\email{vbuendiar@onsager.ugr.es}
\affiliation{Department of Computer Science, University of Tübingen, Tübingen, Germany}
\affiliation{Max Planck Institute for Biological Cybernetics, Tübingen, Germany}
\affiliation{Department of Computing Sciences, Bocconi University, Milan, Italy}

\date{\today}

\begin{abstract}

The celebrated Ott-Antonsen ansatz for coupled oscillators provides a useful framework to work with deterministic systems in the thermodynamic limit, but remains just an approximation for stochastic models. In this paper, I construct a general mesoscopic description of finite-sized populations of stochastic coupled oscillators and apply it to study the stochastic Kuramoto model. From such a mesoscopic description it is possible to obtain the natural, multiplicative fluctuations of the oscillator ensemble. The analysis allows one to derive highly accurate, closed expressions for the stochastic Kuramoto model's order parameter for the first time. Moreover, it is possible to get novel insights into the system's fluctuations and the synchronization transition's critical exponents which were inaccessible before. 
\end{abstract}

                              
\maketitle


\emph{Introduction.} Many interacting systems in Nature exhibit both stochastic and oscillatory behaviour, from physics to living organisms. Thus the interest to understand, model and accurately predict complex systems has led to great advances in the theory of coupled oscillators. These advances have found applications for biological rhythms \cite{mirollo_synchronization_1990}, arrays of coupled Josephson functions \cite{wiesenfeld_frequency_1998}, the study of the power grid \cite{totz_control_2020} or nanoelectronics \cite{matheny_exotic_2019}. In the last years, coupled oscillator models have played an important role in neuroscience, from the study of dynamics \cite{montbrio2015PRX, montbrio_exact_2020, ferrara_population_2023}  to whole-brain models \cite{pang2021N, odor2019SR, daffertshofer2018Na}. While many different approaches have been studied in this context, the Kuramoto model still stands out for its simplicity and explanative power. 

One of the most important theoretical advances to understanding oscillators' dynamics is due to Ott and Antonsen, who introduced an ansatz which provides exact solutions to the classical Kuramoto model \cite{ott2008C, ott2009CIJNS}. Moreover, the Ott-Antonsen ansatz can be used to obtain equations describing the macroscopic behaviour of a large family of non-linear coupled oscillators in the thermodynamic limit. For instance, Ott-Antonsen theory has been successfully applied to the forced Kuramoto model \cite{childs2008CIJNS} or quadratic integrate-and-fire neuronal oscillators \cite{montbrio2015PRX, laing2018JMN}. 

Despite its success, there are limitations. Two of the most important are that the ansatz (i) is only valid in the thermodynamic limit, and (ii) in the presence of stochastic noise, it constitutes only an approximation \cite{goldobin2018CIJNS, tyulkina2018PRL}. The first problem can be addressed using other ansatzs, such as the Watanabe-Strogatz one\cite{watanabe1993PRL}. The second can be mitigated by considering circular cumulants\cite{tyulkina2018PRL, goldobin2018CIJNS, goldobin2021C}. However, one must note that the Watanabe-Strogatz ansatz is only valid for deterministic oscillators, and the use of cumulants does not always allow closed-form analytics. In addition to that, none of these strategies gives information about the nature of the fluctuations of the order parameter themselves, an essential ingredient to study the properties of phase transitions from the point of view of statistical mechanics \cite{binney2001, hohenberg1977RMP}. 

The lack of noise in low-dimensional reductions is not just a theoretical problem, but a practical limitation. For example, in many neuroscience applications, populations of neurons are characterized by their Kuramoto order parameter $R_1$, which measures the neurons' degree of synchrony within a population. Then, the populations are coupled and supplemented with additive white noise \cite{ponce-alvarez2015PCB, cabral2017N, daffertshofer2018Na}. This approach is only valid for weakly-synchronized systems, since in the synchronous regime the noise can push the system towards unphysical values of the Kuramoto order parameter, $R_1 > 1$. For such effective models, one would expect to have an absorbing barrier, i.e., vanishing fluctuations at $R_1 = 1$, which is currently missing.

In this paper, I construct a novel mesoscopic theory for oscillators that explicitly accounts for stochastic fluctuations. Such theory recovers the exact correlation structure of the Kuramoto-Daido order parameters' fluctuations, leading to new insights and capturing finite-size effects extremely accurately. The paper is structured as follows: first, I introduce the main ideas and derive fluctuating equations for the mean-field amplitude and phase. Then, I apply the theory to the stochastic Kuramoto model. In this analysis, I derive closed solutions for the order parameter of the stochastic Kuramoto model for the first time. Finally, I use the theory to further study the synchronization transition's universality class.


\emph{Theory.} Let me start by considering a general system of $N$ fully coupled, identical oscillators, subject to uncorrelated Gaussian white noise,

\begin{equation}
    \dot \phi _i = \omega + F\left[\phi_i, \left\{ Z_k \right\} \right] + \sigma \eta_i(t),
\end{equation}

where $F\left[\phi_i, \left\{ Z_k \right\}\right]$ contains non-linearities and also the couplings to the mean-field, which are encoded using the Kuramoto-Daido parameters $Z_k(t) =N^{-1} \sum_j e^{i k \phi_j(t)}$, and $\eta_i$ are uncorrelated Gaussian white noises, $\langle \eta_i(t) \eta_j(t') \rangle = \delta(t-t') \delta_{ij}$. For example, the mean-field Kuramoto model is given by $F\left[ Z_1 \right] = \operatorname{Im}(J Z_1 e^{-i\phi_i}) = J N^{-1} \sum_j \sin(\phi_j - \phi_i)$, being $J$ the coupling between oscillators. 

Usually, one assumes the thermodynamic limit $N\to+\infty$ and writes the continuity equation for the density of oscillators between $\phi$ and $\phi+d\phi$ \cite{ott2008C}. In the stochastic case, the Fokker-Planck equation plays the role of the continuity equation \cite{acebron2005RMP, tyulkina2018PRL}. The next step is to expand the density in Fourier series and study the evolution of the Fourier coefficients. When the frequencies are homogeneous, the coefficients are $Z_k$, the Kuramoto-Daido parameters. Finally, the Ott-Antonsen ansatz assumes that $Z_k = Z_1^k$, yielding a single equation for the complex number $Z_1$.

The starting point of the mesoscopic theory is the Dean-Kawasaki equation \cite{dean1996JPAMG}. Its derivation is standard (see Appendix A) and leads to a Langevin equation for the evolution of the oscillators' density that generalizes the Fokker-Planck equation,

\begin{align}
   \frac{\partial \rho(\phi, t)}{\partial t} &= -\frac{\partial}{\partial \phi} \left [ \rho(\phi, t) (\omega + F\left[\phi, \left\{ Z_k \right\}\right]) \right] + \nonumber \\ 
   +& \frac{\sigma^2}{2} \frac{\partial^2 \rho(\phi, t)}{\partial \phi^2} + \frac{\sigma}{\sqrt{N}} \frac{\partial}{\partial \phi} \left[ \sqrt{\rho(\phi, t) }\eta(\phi, t) \right] 
   \label{eq:dk-density}
\end{align}

where  $\eta(\phi, t)$ is an uncorrelated Gaussian white noise field\footnote{The noise field $\eta(\phi, t)$ does depend on the angle value, while the individual noises $\eta_j(t)$ are just a function of time. Despite the similarity of notation, which reflects that the term $\sqrt\rho \eta$ emerges from $\eta_j$, both terms are different.}, $\langle \eta(\phi, t)\eta(\phi', t')\rangle =\delta(\phi-\phi')\delta(t-t') $. The noise term appears inside a derivative, ensuring the conservation of density, and its statistics match those of the microscopic system. The equation is to be interpreted in the Itô sense\cite{dean1996JPAMG}.  It is essential to remark that the deterministic part of eq.(\ref{eq:dk-density}) coincides exactly with that of the Fokker-Planck. The Kuramoto-Daido parameters can be obtained from their definition,

\begin{equation}
    Z_k (t) = \int_0 ^{2\pi} d\phi \rho(\phi, t) e^{ik\phi}  \equiv x_k + i y_k \equiv R_k e^{i\psi_k }.
    \label{eq:kd-order}
\end{equation}

Deriving eq. (\ref{eq:dk-density}) with respect to time, inserting eq.~(\ref{eq:kd-order}), and integrating by parts, one finds the time evolution for every $Z_k$, 

\begin{equation}
    \dot Z_k (t) = G_k[\left\{ Z_k \right\}] + \xi_k(t).
\end{equation}

In this expression, $G_k[\left\{ Z_k \right\}]$ represents the terms obtained from the deterministic part. Hence, these coincide with the ones obtained from the Fokker-Planck equation. Therefore, the novelty here lies in the noise term. One can show (see Appendix B) that the integration by parts yields 

\begin{equation}
    \xi_{k}\left(t\right)= -\frac{k\sigma i}{\sqrt{N}}\int d\phi e^{ik\phi}\sqrt{\rho\left(\phi,t\right)}\eta\left(\phi,t\right).
\end{equation}

$\xi_k(t)$ has zero mean, since $\langle \eta(\phi, t) \rangle = 0$, but its correlation matrix needs to be obtained by decomposing it into real and imaginary parts, $\xi_k = \xi_{k,x} + i \xi_{k,y}$, giving

\begin{subequations}
\begin{align}
    \left\langle \xi_{k,x}(t)\xi_{k',x}(t')\right\rangle  =& \frac{kk'\sigma^{2}}{2N}\delta\left(t-t'\right)\left(x_{k-k'}-x_{k+k'}\right), \\
    \left\langle \xi_{k,y}(t)\xi_{k',x}(t')\right\rangle  =& \frac{kk'\sigma^{2}}{2N}\delta\left(t-t'\right)\left(y_{k-k'}-y_{k+k'}\right), \\
    \left\langle \xi_{k,y}(t)\xi_{k',y}(t')\right\rangle  =& \frac{kk'\sigma^{2}}{2N}\delta\left(t-t'\right)\left(x_{k-k'}+x_{k+k'}\right).
\end{align}
\label{eq:correlations}
\end{subequations}

\begin{figure}[!htpb]
\includegraphics[width=\columnwidth]{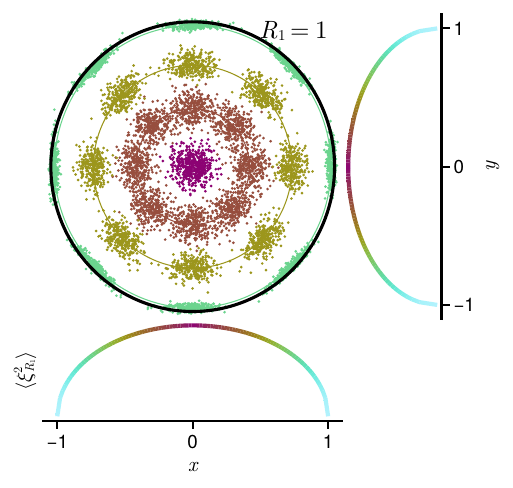}
\caption{ \textbf{Order parameters' correlation structure generates an absorbing barrier at full synchronization}. Random Gaussian variables are generated from the correlation matrix in eqs.~(\ref{eq:correlations}) for several values of $Z_1$. For illustration purposes, the matrix is simplified considering the Ott-Antonsen ansatz, so it depends only on $Z_1$ and it is model-agnostic. Fluctuations close to $R_1=1$ become tangent to the circle. Thus, the intensity of the stochastic fluctuations in the $R_1$ direction vanishes at $R_1 = 1$, as shown in the surrounding axes} \label{fig:absorbingnoise}
\end{figure}

The correlation between the real and imaginary coordinates $x_k$ and $y_k$ of the Kuramoto-Daido parameters  (defined in eq. (\ref{eq:kd-order})) prevents the system from reaching values $R_k > 1$ for all $k$, effectively generating an absorbing barrier at $R_k=1$, as shown in Fig.~\ref{fig:absorbingnoise}. This can be checked by changing variables to amplitude and angle variables. The algebra is simple, but rather tedious (and can be found in Appendix C). After the change to polar coordinates, one is faced with the following Langevin equations for the Kuramoto-Daido amplitudes and phases, 

\begin{subequations}
\begin{align}
    \dot R_k &= f\left(\{ R_k, \psi_k \}\right) + \frac{\Lambda_{R,k}\left(\{ R_k, \psi_k \}\right)}{N} + \xi_{k, R}(t)\\
    \dot \psi_k &= g\left(\{ R_k, \psi_k \}\right) + \frac{\Lambda_{\psi,k}\left(\{ R_k, \psi_k \}\right)}{N} + \xi_{k \psi}(t)
\end{align}
\label{eqs:general-system}
\end{subequations}

where the functions $f$ and $g$ give the deterministic behaviour obtained from the usual Fokker-Planck description, and  $\Lambda_{R,k}, \Lambda_{\psi,k}$ are extra deterministic drifts arising from the application of Itô's Lemma. I would like to highlight that such terms emerge because of the multiplicative nature of the noise, and would not present for additive one. The noises $\xi_{k,R}$ and $\xi_{k,\psi}$ are white and Gaussian, but correlated. The full form of the noise correlations and deterministic drifts is written in Appendix C. 

As an illustrative example, let me write explicitly the correlations for the amplitude $R_k$ at equal times,
\begin{align}
     \left\langle \xi_{k,R}\xi_{k',R}\right\rangle =&\frac{kk'\sigma^{2}}{2N} 
    \left[R_{k-k'}\cos\left(\psi_{k'-k}-\left(\psi_{k'}-\psi_{k}\right)\right)-\right. \nonumber \\
    -& \left. R_{k+k'} \cos\left(\psi_{k'+k}-\left(\psi_{k'}+\psi_{k}\right)\right)\right].
    \label{eq:excorrRk}
\end{align}

When oscillators are fully synchronous, then $Z_k = e^{ik\phi}$, so $R_k = 1$ and $\psi_k = k \psi_1$. At this point the oscillator's noise intensity vanishes, not allowing it to cross the $R_k = 1$ barrier. Additionally, angle and amplitude decorrelate when the system becomes fully synchronized. 

This is the first main result of the paper: when the microscopic oscillators are driven by white, uncorrelated noise, the system's Kuramoto-Daido parameters are affected by non-trivial multiplicative fluctuations and a deterministic drift of order $1/N$. Observe that the result obtained here is agnostic of the deterministic oscillator model employed: I only have added an additional fluctuating term to the Fokker-Planck equation --the Dean-Kawasaki formalism-- and worked out the effect of the fluctuations in the order parameters. Thus, they can supplement already existing models just by adding the new drift and fluctuation terms. 

It is worth remarking that the theory can be easily extended to multiplicative noise for the individual oscillators, as well as heterogeneity in the frequencies, as shown in Appendix B.


\emph{Application to the stochastic Kuramoto model.} From now on I will focus on the stochastic Kuramoto model to showcase the advantages of the theory.  First, I analyse how the information on the system's fluctuations structure can help us to gain new conceptual insights that were inaccessible before. 

The Kuramoto model is symmetric under rotations, i.e., any change $\phi_i \to \phi_i + \alpha$ leaves the model invariant. However, the collective phase $\psi_1$ appears explicitly correlated to higher-order collective phases by terms such as eq. (\ref{eq:excorrRk}). This means that the intensity of fluctuations felt by the oscillators could depend on their (collective) position on the circle, breaking the symmetry. I hypothesize that the fluctuation strength should be also symmetric under rotations. The simplest ansatz to guarantee such symmetry is  $\psi_k = k\psi_1$, which cancels all angular dependency in fluctuations. This corresponds to the classical Ott-Antonsen ansatz, when is applied to the phases exclusively. 

Then, I only need to set $\psi_k = k\psi_1$. Just this assumption greatly simplifies the stochastic Kuramoto model, since all phases can depend on $\psi_1$ and all the amplitude equations and their correlations decouple from the phases. Thus, it is sufficient to study the system 

\begin{subequations}
\begin{flalign}
        \dot{R}_{k}= \frac{J}{2}kR_1 \left(R_{k-1}-R_{k+1}\right) -\frac{\sigma^2 k^2}{2}R_k + &&\nonumber\\
     +\frac{k^2 \sigma^2}{2NR_{k}}\left(1+R_{k}^{2}\right) + \xi_{R, k}(t), &&\\
    \langle  \xi_{R, k}(t) \xi_{R, k'}(t) \rangle =\frac{kk'\sigma^2}{2N} (R_{k-k'} - R_{k+k'}),&&
\end{flalign}
\label{eq:simple_kuramoto}
\end{subequations}

with $\langle  \xi_{R, k} \rangle = 0$. The deterministic drift $\Lambda_{R,k}$ is explicitly written. 
Eqs. (\ref{eq:simple_kuramoto}) are our main object of study, and represent all the order parameters of the finite-size stochastic Kuramoto model and their fluctuations.

Let me focus first on the thermodynamic limit, neglecting all $1/N$ terms and stochastic fluctuations for now. At the stationary state, we have $\dot R_k = 0$, which is a non-linear recurrence for $R_k$. It admits a closed solution\footnote{Can be directly obtained with the aid of computational algebra system. I used WolframEngine.} in terms of modified Bessel functions (see Appendix D). 

\begin{figure*}[!htpb]
\includegraphics[width=\textwidth]{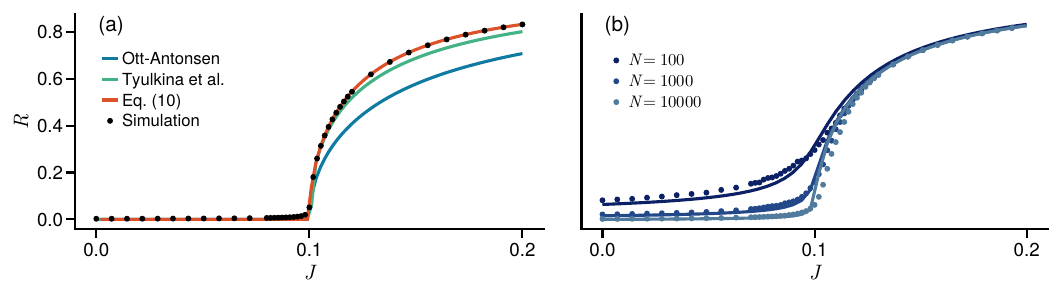}
\caption{ \textbf{Order parameter of the stochastic Kuramoto model.} (a) Simulations of a stochastic Kuramoto model with $N=10^5$ oscillators (black dots) compared to several theories, including eq. (\ref{eq:order-6th}) (red line). Such an analytic solution improves previous results (blue, green lines) in the synchronized phase. (b) Stationary solutions of eq.(\ref{eq:simple_kuramoto}) capture the finite size residual synchronization in the incoherent state. Simulations of the Kuramoto model were done with $\sigma^2 = 0.1$.} \label{fig:order_parameter}
\end{figure*}

Such solution does not tell us the value of the Kuramoto order parameter, $R_1$, which serves as a initial condition to the recurrence. A way to obtain it is to use a closure scheme: setting $R_{k\geq k^*} = 0$ for a certain $k^*$ will suffice. Observe that this is conceptually no different from integrating the original stochastic differential equations eqs.~(\ref{eq:simple_kuramoto}) in a computer, integrating a finite number of variables, and assuming a condition (closure) to cut the infinite hierarchy. The great advantage is that now we just need to find a root of an algebraic expression, instead of integrating $2k^*$ differential equations. 

Not only the algebraic equation for the stationary state can be solved numerically, but it also admits closed solutions up at least up to $k^*=6$. Setting $R_{6} = 0$, the Kuramoto order parameter reads 

\begin{equation}
    R_1=\frac{\sqrt{2}\sigma^{2}}{J}\sqrt{\frac{24\sigma^{2}-9J-\sqrt{51J^{2}-132J\sigma^{2}+306\sigma^{4}}}{J-9\sigma^{2}}},
    \label{eq:order-6th}
\end{equation}

which is valid only when the square root is positive. The critical transition takes place when the square root becomes negative, at $J_c=\sigma^2$, as originally predicted by the Ott-Antonsen ansatz. 

Closure methods have been used in the past for deterministic ansatzs. For instance, Tyulkina \emph{et al.} used a similar approach, writing dynamical equations for the first two circular cumulants, $Z_1$ and $C_2 = Z_2 - Z_1 ^2$, while fixing the third\cite{tyulkina2018PRL}. However, the authors considered both amplitudes and phases, so the system needs to be numerically integrated to obtain the order parameter.
I compare Tyulkina \emph{et al.}'s previous results to eq. ({\ref{eq:order-6th}}) in figure \ref{fig:order_parameter}a. Despite the great improvement of Tyulkina \emph{et al.} over the Ott-Antonsen ansatz, eq.~(\ref{eq:order-6th}) provides a significantly better result. Up to my knowledge, this is the first time such an accurate closed expression has been derived for the stochastic Kuramoto model.

So far, I have analysed the thermodynamic limit of eq. (\ref{eq:simple_kuramoto}). But the new theory also allows studying finite-size effects in a simple setting. One can neglect the fluctuations in eq. (\ref{eq:simple_kuramoto}), but keeping the $\mathcal{O}(1/N)$ deterministic drifts. The resulting stationary values are shown in Fig.~\ref{fig:order_parameter}b against simulations of the Kuramoto model for different system sizes. Observe that the theory predicts very accurately the residual synchronization observed in the asynchronous regime, even for small system sizes, down to $N=100$ oscillators. 
Moreover, considering the Ott-Antonsen ansatz, it is possible to write down a closed formula for $R_1(N)$ in the stationary state (see Appendix D). The formula fails for the synchronous regime, as expected by the Ott-Antonsen ansatz, but it is able to match the simulations in the asynchronous one, revealing that the ansatz is attracting even at small sizes in the incoherent region. As far as I am concerned this is the first time that such kind of results have been reported for the Kuramoto model.


\emph{Fluctuations and critical properties.} The last step is to finally study the effect of fluctuations. I have shown that the theory is able to predict order parameters, including finite-size effects, extremely accurately. Now, I integrate the full set of Langevin eqs. (\ref{eq:simple_kuramoto}). Due to the system's complicated correlation structure, the numerical simulation is not straightforward and additional details are provided in Appendix E. 

\begin{figure}[!htpb]
\includegraphics[width=\columnwidth]{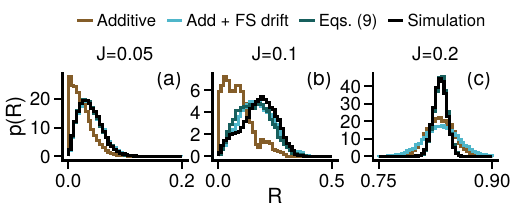}%
\caption{ \textbf{Distribution of the Kuramoto order parameter.} Distribution of $R_1$ for simulations of a stochastic Kuramoto model with $N=1000$ oscillators (black line), additive noise (brow line), additive noise plus the finite-size drifts (light blue) and full integration of eqs. (\ref{eq:simple_kuramoto}) (green line).  Simulations are done for three coupling values, including the critical point $J_c=\sigma^2=0.1$. Relaxation time, $t_r = 7000$.} \label{fig:r-distributions}
\end{figure}

In Fig.~\ref{fig:r-distributions}, I show that the mesoscopic model recovers the entire distribution of the Kuramoto order parameters for any coupling, even at the critical point and for small system sizes. These results are compared to the addition of white noise of variance $\sigma^2/(2N)$ to the deterministic order parameter equations. The additive noise model fails to reproduce the simulations at any point. However, adding the finite-size drift to the additive noise is sufficient to recover the distributions in the asynchronous case, up to the critical point. In the synchronous regime, close to the absorbing barrier, the full structure of correlations is required to match the simulations, since the additive noise greatly overestimates the distribution's variance.

Finally, I study the critical properties of the synchronization transition. This was impossible in previous ansatz reductions since they did not provide information on the order parameter's fluctuations. For simplicity, let me consider the Ott-Antonsen reduction of eq.~(\ref{eq:simple_kuramoto}),

\begin{equation}
    \dot{R_1}=\frac{R_1}{2}\left(J-\sigma^{2}\right) -\frac{JR_1^{3}}{2} +\frac{\sigma^{2}\left(1+R_1 ^2\right)}{2NR_1} + \sigma\sqrt{\frac{1-R_1^{2}}{2N}}\xi(t),
    \label{eq:stochastic-oa-kuramoto}
\end{equation}

Since critical properties are only well-defined in the thermodynamic limit, I will neglect the finite-size drift --of order $\mathcal{O}(1/N)$-- but keep the stochastic term, essential to criticality. Moreover, at criticality I still have $R_1 \ll 1$, meaning that the noise term can be Taylor expanded to $\mathcal{O}(R_1)$, leaving 

\begin{equation}
    \dot{R_1}=\frac{1}{2}\left[\left(J-\sigma^{2}\right)R_1-J R_1^{3}\right] + \sqrt{\frac{\sigma^2}{2N}}\xi(t).
    \label{eq:critical-kuramoto}
\end{equation}

Note that equation (\ref{eq:critical-kuramoto}) is the usual Stuart-Landau model with additive white noise. This Langevin equation is identical to the Hohenberg-Halperin model A, and hence it is expected to display the usual mean-field critical exponents\cite{hohenberg1977RMP}. An interesting consequence is the fact that the critical exponent $\gamma$ --defined as $\chi\sim \varepsilon^\gamma$, being $\chi=\langle R_1 ^2 \rangle - \langle R_1 \rangle^2$ the susceptibility and $\varepsilon=(J-J_c)/J_c$ the reduced distance to criticality-- is expected to be 1. The exponent $\gamma$ has been shown to be particularly difficult to obtain in the case with random frequencies\cite{hong2015PRE}. Since the deterministic part of the Ott-Antonsen ansatz for a Kuramoto model with Lorentzian-distributed frequencies is identical to the stochastic one\cite{buendia2021PRR}, and the ansatz becomes exact in the limit $\sigma\to 0$, I believe the argument offers strong support to $\gamma = \gamma' = 1$ exactly, as argued by Hong \emph{et al} \cite{hong2015PRE}.  

\begin{figure}[!htpb]
\includegraphics[width=\columnwidth]{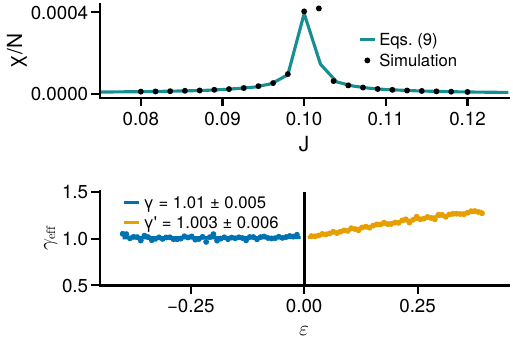}
\caption{ \textbf{Estimation of the $\gamma$ exponent} (a) Susceptibility as a function of the coupling parameter $J$ for a Kuramoto model with $N=10^5$ oscillators compared to eqs. (\ref{eq:simple_kuramoto}) with additive noise. Susceptibility is estimated as the variance of $R$ across 100 trajectories. Simulation parameters were as in Fig.~\ref{fig:r-distributions}. (b) The effective exponent as a function of the reduced distance to criticality $\varepsilon = (J - J_c) / J_c$, obtained from the variance of $R$, averaged over 6000 simulations. Extrapolation is linear in the asynchronous regime, but quadratic in the synchronous one. Both lines converge to the same value of $\gamma$ at $J=J_c$.} \label{fig:gamma-exponent}
\end{figure}

However, as the noise intensity $\sigma$ grows, the Ott-Antonsen approximation becomes inaccurate on the supercritical side of the transition (as we saw in Fig.~\ref{fig:order_parameter}). Thus, for the stochastic version of the Kuramoto model, $\gamma = 1$, but potentially one could have $\gamma' = 1 + H(\sigma)$, with $H(0)=0$. Thus, I estimated the critical exponent $\gamma$ following Hong \emph{et al.}, by computing the off-critical gamma exponents using eqs. ~(\ref{eq:simple_kuramoto}) with additive noise and finite-size drift. The results are shown in Fig. \ref{fig:gamma-exponent}, which suggests that $\gamma' = 1$. This is justified since the distribution of $R_1$ is well-captured just by additive noise close to the critical point, only failing for well-synchronized values (as seen in Fig.~\ref{fig:r-distributions}). 

\emph{Discussion.} In this paper, I have presented a mesoscopic model able to account for the fluctuations in the Kuramoto-Daido order parameters when the coupled oscillators are subject to stochastic noise. In this case, it was well-known that the Ott-Antonsen ansatz is only approximate up to $\mathcal{O}(\sigma^2)$ \cite{goldobin_mean-field_2021} and some solutions, such as circular cumulants \cite{tyulkina2018PRL} and more recently pseudo-cumulants \cite{goldobin_reduction_2021} have been proposed. However, all previous solutions deal with the average value of the Kuramoto-Daido parameters in the thermodynamic limit, losing all information about stochastic fluctuations and finite size. The novel approach presented here is able to recover the entire distribution of the order parameter. Even when just average values are considered, the correlation structure hinted a way to obtain a closed expression for the stochastic Kuramoto model for the first time. The finite-size driftings also allow for further corrections that, up to my knowledge, have not been presented before and open the door to novel analyses. 

But this is not the end of the story. So far, I have focused only on the stochastic Kuramoto model for simplicity. I have shown that close to the critical transition, additive noise is good enough to recover most of the phenomenology. However, other oscillator models have transitions happening at high synchrony. For instance, stochastic oscillators of the form $\dot\phi = \omega + a\sin\phi$ can be coupled, leading to complex behaviour, including collective saddle-node on invariant circle (SNIC) and co-dimension 2 bifurcations \cite{childs2008CIJNS, buendia2021PRR}. In a collective SNIC transition with multiplicative noise, fluctuations will be highly suppressed, since $R_1 \simeq 1$, and noise in the transition will be purely multiplicative, leading to potentially new universality classes. These could not be studied before in such detail, as the exact form of the fluctuations was missing. Furthermore, such oscillators are also commonplace in neuroscience, taking place as quadratic integrate-and-fire neurons \cite{laing2018JMN, goldobin_mean-field_2021}. We previously studied the critical properties of such oscillators in \cite{buendia2021PRR}, but an analytical treatment was impossible. Another study focusing on critical properties of mesoscopic SNIC bifurcations shall be soon published elsewhere. 

Finally, I would like to comment briefly on the relationship between this method and the circular cumulant approach \cite{tyulkina2018PRL, goldobin2018CIJNS} which has been successfully applied not only for the Kuramoto model but also for quadratic-and-fire neurons \cite{goldobin_mean-field_2021, goldobin_reduction_2021, dolmatova_circular_2023}. The method is fully compatible with the ideas presented here. In fact, the stochastic equations for the order parameters could be rewritten as equations for the circular cumulants (or pseudocumulants). Since solving for cumulants is much more precise than working with raw moments, this has the potential to improve the results presented here. So far, in this paper the mesoscopic method outperforms the circular cumulants due to the simplification of the phases --which was hinted by the structure of fluctuations-- but I believe that combining both ideas will lead to fruitful results in the short term. 

In summary, I hope the theory presented here will help to deepen our understanding of the dynamics of coupled oscillators subject to stochastic fluctuations, leading to new results on critical bifurcations, and finite-size effects, which were not attainable to analytical methods before.


\emph{Acknowledgements.} This project was fully supported by a Sofja Kovalevskaya Award during my postdoc at the Univ. T\"ubingen and the MPI of Biological Cybernetics; Now I acknowledge funding by the NextGenerationEU, in the framework of the FAIR - Future Artificial Intelligence Research project (FAIR PE00000013 – CUP B43C22000800006). Simulations were done at the PROTEUS cluster, funded by the Spanish Ministry and Agencia Estatal de investigación (AEI) through Project I+D+i Ref. PID2020-113681GB-I00, financed by MICIN/AEI/10.13039/501100011033. I would like to thank Rub\'en Hurtado, Matteo Sirecci, and Giorgio Nicoletti for insightful discussions on oscillators. I would like to acknowledge also Ana Paula Millán, Miguel A. Muñoz and Nicolas Brunel for invaluable feedback on the manuscript. Finally, I thank Anna Levina, for her great support and supervision during my postdoc in T\"ubingen.


\begin{thebibliography}{31}%
\makeatletter
\providecommand \@ifxundefined [1]{%
 \@ifx{#1\undefined}
}%
\providecommand \@ifnum [1]{%
 \ifnum #1\expandafter \@firstoftwo
 \else \expandafter \@secondoftwo
 \fi
}%
\providecommand \@ifx [1]{%
 \ifx #1\expandafter \@firstoftwo
 \else \expandafter \@secondoftwo
 \fi
}%
\providecommand \natexlab [1]{#1}%
\providecommand \enquote  [1]{``#1''}%
\providecommand \bibnamefont  [1]{#1}%
\providecommand \bibfnamefont [1]{#1}%
\providecommand \citenamefont [1]{#1}%
\providecommand \href@noop [0]{\@secondoftwo}%
\providecommand \href [0]{\begingroup \@sanitize@url \@href}%
\providecommand \@href[1]{\@@startlink{#1}\@@href}%
\providecommand \@@href[1]{\endgroup#1\@@endlink}%
\providecommand \@sanitize@url [0]{\catcode `\\12\catcode `\$12\catcode
  `\&12\catcode `\#12\catcode `\^12\catcode `\_12\catcode `\%12\relax}%
\providecommand \@@startlink[1]{}%
\providecommand \@@endlink[0]{}%
\providecommand \url  [0]{\begingroup\@sanitize@url \@url }%
\providecommand \@url [1]{\endgroup\@href {#1}{\urlprefix }}%
\providecommand \urlprefix  [0]{URL }%
\providecommand \Eprint [0]{\href }%
\providecommand \doibase [0]{https://doi.org/}%
\providecommand \selectlanguage [0]{\@gobble}%
\providecommand \bibinfo  [0]{\@secondoftwo}%
\providecommand \bibfield  [0]{\@secondoftwo}%
\providecommand \translation [1]{[#1]}%
\providecommand \BibitemOpen [0]{}%
\providecommand \bibitemStop [0]{}%
\providecommand \bibitemNoStop [0]{.\EOS\space}%
\providecommand \EOS [0]{\spacefactor3000\relax}%
\providecommand \BibitemShut  [1]{\csname bibitem#1\endcsname}%
\let\auto@bib@innerbib\@empty
\bibitem [{\citenamefont {Mirollo}\ and\ \citenamefont
  {Strogatz}(1990)}]{mirollo_synchronization_1990}%
  \BibitemOpen
  \bibfield  {author} {\bibinfo {author} {\bibfnamefont {R.~E.}\ \bibnamefont
  {Mirollo}}\ and\ \bibinfo {author} {\bibfnamefont {S.~H.}\ \bibnamefont
  {Strogatz}},\ }\bibfield  {title} {\bibinfo {title} {Synchronization of
  {Pulse}-{Coupled} {Biological} {Oscillators}},\ }\href
  {https://doi.org/10.1137/0150098} {\bibfield  {journal} {\bibinfo  {journal}
  {SIAM Journal on Applied Mathematics}\ }\textbf {\bibinfo {volume} {50}},\
  \bibinfo {pages} {1645} (\bibinfo {year} {1990})},\ \bibinfo {note}
  {publisher: Society for Industrial and Applied Mathematics}\BibitemShut
  {NoStop}%
\bibitem [{\citenamefont {Wiesenfeld}\ \emph {et~al.}(1998)\citenamefont
  {Wiesenfeld}, \citenamefont {Colet},\ and\ \citenamefont
  {Strogatz}}]{wiesenfeld_frequency_1998}%
  \BibitemOpen
  \bibfield  {author} {\bibinfo {author} {\bibfnamefont {K.}~\bibnamefont
  {Wiesenfeld}}, \bibinfo {author} {\bibfnamefont {P.}~\bibnamefont {Colet}},\
  and\ \bibinfo {author} {\bibfnamefont {S.~H.}\ \bibnamefont {Strogatz}},\
  }\bibfield  {title} {\bibinfo {title} {Frequency locking in {Josephson}
  arrays: {Connection} with the {Kuramoto} model},\ }\href
  {https://doi.org/10.1103/PhysRevE.57.1563} {\bibfield  {journal} {\bibinfo
  {journal} {Physical Review E}\ }\textbf {\bibinfo {volume} {57}},\ \bibinfo
  {pages} {1563} (\bibinfo {year} {1998})},\ \bibinfo {note} {publisher:
  American Physical Society}\BibitemShut {NoStop}%
\bibitem [{\citenamefont {Totz}\ \emph {et~al.}(2020)\citenamefont {Totz},
  \citenamefont {Olmi},\ and\ \citenamefont {Schöll}}]{totz_control_2020}%
  \BibitemOpen
  \bibfield  {author} {\bibinfo {author} {\bibfnamefont {C.~H.}\ \bibnamefont
  {Totz}}, \bibinfo {author} {\bibfnamefont {S.}~\bibnamefont {Olmi}},\ and\
  \bibinfo {author} {\bibfnamefont {E.}~\bibnamefont {Schöll}},\ }\bibfield
  {title} {\bibinfo {title} {Control of synchronization in two-layer power
  grids},\ }\href {https://doi.org/10.1103/PhysRevE.102.022311} {\bibfield
  {journal} {\bibinfo  {journal} {Physical Review E}\ }\textbf {\bibinfo
  {volume} {102}},\ \bibinfo {pages} {022311} (\bibinfo {year} {2020})},\
  \bibinfo {note} {publisher: American Physical Society}\BibitemShut {NoStop}%
\bibitem [{\citenamefont {Matheny}\ \emph {et~al.}(2019)\citenamefont
  {Matheny}, \citenamefont {Emenheiser}, \citenamefont {Fon}, \citenamefont
  {Chapman}, \citenamefont {Salova}, \citenamefont {Rohden}, \citenamefont
  {Li}, \citenamefont {Hudoba~de Badyn}, \citenamefont {Pósfai}, \citenamefont
  {Duenas-Osorio}, \citenamefont {Mesbahi}, \citenamefont {Crutchfield},
  \citenamefont {Cross}, \citenamefont {D’Souza},\ and\ \citenamefont
  {Roukes}}]{matheny_exotic_2019}%
  \BibitemOpen
  \bibfield  {author} {\bibinfo {author} {\bibfnamefont {M.~H.}\ \bibnamefont
  {Matheny}}, \bibinfo {author} {\bibfnamefont {J.}~\bibnamefont {Emenheiser}},
  \bibinfo {author} {\bibfnamefont {W.}~\bibnamefont {Fon}}, \bibinfo {author}
  {\bibfnamefont {A.}~\bibnamefont {Chapman}}, \bibinfo {author} {\bibfnamefont
  {A.}~\bibnamefont {Salova}}, \bibinfo {author} {\bibfnamefont
  {M.}~\bibnamefont {Rohden}}, \bibinfo {author} {\bibfnamefont
  {J.}~\bibnamefont {Li}}, \bibinfo {author} {\bibfnamefont {M.}~\bibnamefont
  {Hudoba~de Badyn}}, \bibinfo {author} {\bibfnamefont {M.}~\bibnamefont
  {Pósfai}}, \bibinfo {author} {\bibfnamefont {L.}~\bibnamefont
  {Duenas-Osorio}}, \bibinfo {author} {\bibfnamefont {M.}~\bibnamefont
  {Mesbahi}}, \bibinfo {author} {\bibfnamefont {J.~P.}\ \bibnamefont
  {Crutchfield}}, \bibinfo {author} {\bibfnamefont {M.~C.}\ \bibnamefont
  {Cross}}, \bibinfo {author} {\bibfnamefont {R.~M.}\ \bibnamefont
  {D’Souza}},\ and\ \bibinfo {author} {\bibfnamefont {M.~L.}\ \bibnamefont
  {Roukes}},\ }\bibfield  {title} {\bibinfo {title} {Exotic states in a simple
  network of nanoelectromechanical oscillators},\ }\href
  {https://doi.org/10.1126/science.aav7932} {\bibfield  {journal} {\bibinfo
  {journal} {Science}\ }\textbf {\bibinfo {volume} {363}},\ \bibinfo {pages}
  {eaav7932} (\bibinfo {year} {2019})},\ \bibinfo {note} {publisher: American
  Association for the Advancement of Science}\BibitemShut {NoStop}%
\bibitem [{\citenamefont {Montbrió}\ \emph {et~al.}(2015)\citenamefont
  {Montbrió}, \citenamefont {Pazó},\ and\ \citenamefont
  {Roxin}}]{montbrio2015PRX}%
  \BibitemOpen
  \bibfield  {author} {\bibinfo {author} {\bibfnamefont {E.}~\bibnamefont
  {Montbrió}}, \bibinfo {author} {\bibfnamefont {D.}~\bibnamefont {Pazó}},\
  and\ \bibinfo {author} {\bibfnamefont {A.}~\bibnamefont {Roxin}},\ }\bibfield
   {title} {\bibinfo {title} {Macroscopic {{Description}} for {{Networks}} of
  {{Spiking Neurons}}},\ }\href {https://doi.org/10.1103/PhysRevX.5.021028}
  {\bibfield  {journal} {\bibinfo  {journal} {Physical Review X}\ }\textbf
  {\bibinfo {volume} {5}},\ \bibinfo {pages} {021028} (\bibinfo {year}
  {2015})}\BibitemShut {NoStop}%
\bibitem [{\citenamefont {Montbrió}\ and\ \citenamefont
  {Pazó}(2020)}]{montbrio_exact_2020}%
  \BibitemOpen
  \bibfield  {author} {\bibinfo {author} {\bibfnamefont {E.}~\bibnamefont
  {Montbrió}}\ and\ \bibinfo {author} {\bibfnamefont {D.}~\bibnamefont
  {Pazó}},\ }\bibfield  {title} {\bibinfo {title} {Exact {Mean}-{Field}
  {Theory} {Explains} the {Dual} {Role} of {Electrical} {Synapses} in
  {Collective} {Synchronization}},\ }\href
  {https://doi.org/10.1103/PhysRevLett.125.248101} {\bibfield  {journal}
  {\bibinfo  {journal} {Physical Review Letters}\ }\textbf {\bibinfo {volume}
  {125}},\ \bibinfo {pages} {248101} (\bibinfo {year} {2020})},\ \bibinfo
  {note} {publisher: American Physical Society}\BibitemShut {NoStop}%
\bibitem [{\citenamefont {Ferrara}\ \emph {et~al.}(2023)\citenamefont
  {Ferrara}, \citenamefont {Angulo-Garcia}, \citenamefont {Torcini},\ and\
  \citenamefont {Olmi}}]{ferrara_population_2023}%
  \BibitemOpen
  \bibfield  {author} {\bibinfo {author} {\bibfnamefont {A.}~\bibnamefont
  {Ferrara}}, \bibinfo {author} {\bibfnamefont {D.}~\bibnamefont
  {Angulo-Garcia}}, \bibinfo {author} {\bibfnamefont {A.}~\bibnamefont
  {Torcini}},\ and\ \bibinfo {author} {\bibfnamefont {S.}~\bibnamefont
  {Olmi}},\ }\bibfield  {title} {\bibinfo {title} {Population spiking and
  bursting in next-generation neural masses with spike-frequency adaptation},\
  }\href {https://doi.org/10.1103/PhysRevE.107.024311} {\bibfield  {journal}
  {\bibinfo  {journal} {Physical Review E}\ }\textbf {\bibinfo {volume}
  {107}},\ \bibinfo {pages} {024311} (\bibinfo {year} {2023})},\ \bibinfo
  {note} {publisher: American Physical Society}\BibitemShut {NoStop}%
\bibitem [{\citenamefont {Pang}\ \emph {et~al.}(2021)\citenamefont {Pang},
  \citenamefont {Gollo},\ and\ \citenamefont {Roberts}}]{pang2021N}%
  \BibitemOpen
  \bibfield  {author} {\bibinfo {author} {\bibfnamefont {J.~C.}\ \bibnamefont
  {Pang}}, \bibinfo {author} {\bibfnamefont {L.~L.}\ \bibnamefont {Gollo}},\
  and\ \bibinfo {author} {\bibfnamefont {J.~A.}\ \bibnamefont {Roberts}},\
  }\bibfield  {title} {\bibinfo {title} {Stochastic synchronization of dynamics
  on the human connectome},\ }\href
  {https://doi.org/10.1016/j.neuroimage.2021.117738} {\bibfield  {journal}
  {\bibinfo  {journal} {NeuroImage}\ }\textbf {\bibinfo {volume} {229}},\
  \bibinfo {pages} {117738} (\bibinfo {year} {2021})}\BibitemShut {NoStop}%
\bibitem [{\citenamefont {Ódor}\ and\ \citenamefont
  {Kelling}(2019)}]{odor2019SR}%
  \BibitemOpen
  \bibfield  {author} {\bibinfo {author} {\bibfnamefont {G.}~\bibnamefont
  {Ódor}}\ and\ \bibinfo {author} {\bibfnamefont {J.}~\bibnamefont
  {Kelling}},\ }\bibfield  {title} {\bibinfo {title} {Critical synchronization
  dynamics of the {{Kuramoto}} model on connectome and small world graphs},\
  }\href {https://doi.org/10.1038/s41598-019-54769-9} {\bibfield  {journal}
  {\bibinfo  {journal} {Scientific Reports}\ }\textbf {\bibinfo {volume} {9}},\
  \bibinfo {pages} {19621} (\bibinfo {year} {2019})}\BibitemShut {NoStop}%
\bibitem [{\citenamefont {Daffertshofer}\ \emph {et~al.}(2018)\citenamefont
  {Daffertshofer}, \citenamefont {Ton}, \citenamefont {Kringelbach},
  \citenamefont {Woolrich},\ and\ \citenamefont {Deco}}]{daffertshofer2018Na}%
  \BibitemOpen
  \bibfield  {author} {\bibinfo {author} {\bibfnamefont {A.}~\bibnamefont
  {Daffertshofer}}, \bibinfo {author} {\bibfnamefont {R.}~\bibnamefont {Ton}},
  \bibinfo {author} {\bibfnamefont {M.~L.}\ \bibnamefont {Kringelbach}},
  \bibinfo {author} {\bibfnamefont {M.}~\bibnamefont {Woolrich}},\ and\
  \bibinfo {author} {\bibfnamefont {G.}~\bibnamefont {Deco}},\ }\bibfield
  {title} {\bibinfo {title} {Distinct criticality of phase and amplitude
  dynamics in the resting brain},\ }\href
  {https://doi.org/10.1016/j.neuroimage.2018.03.002} {\bibfield  {journal}
  {\bibinfo  {journal} {NeuroImage}\ }\bibinfo {series} {Brain {{Connectivity
  Dynamics}}},\ \textbf {\bibinfo {volume} {180}},\ \bibinfo {pages} {442}
  (\bibinfo {year} {2018})}\BibitemShut {NoStop}%
\bibitem [{\citenamefont {Ott}\ and\ \citenamefont
  {Antonsen}(2008)}]{ott2008C}%
  \BibitemOpen
  \bibfield  {author} {\bibinfo {author} {\bibfnamefont {E.}~\bibnamefont
  {Ott}}\ and\ \bibinfo {author} {\bibfnamefont {T.~M.}\ \bibnamefont
  {Antonsen}},\ }\bibfield  {title} {\bibinfo {title} {Low dimensional behavior
  of large systems of globally coupled oscillators},\ }\href
  {https://doi.org/10.1063/1.2930766} {\bibfield  {journal} {\bibinfo
  {journal} {Chaos: An Interdisciplinary Journal of Nonlinear Science}\
  }\textbf {\bibinfo {volume} {18}},\ \bibinfo {pages} {037113} (\bibinfo
  {year} {2008})}\BibitemShut {NoStop}%
\bibitem [{\citenamefont {Ott}\ and\ \citenamefont
  {Antonsen}(2009)}]{ott2009CIJNS}%
  \BibitemOpen
  \bibfield  {author} {\bibinfo {author} {\bibfnamefont {E.}~\bibnamefont
  {Ott}}\ and\ \bibinfo {author} {\bibfnamefont {T.~M.}\ \bibnamefont
  {Antonsen}},\ }\bibfield  {title} {\bibinfo {title} {Long time evolution of
  phase oscillator systems},\ }\href@noop {} {\bibfield  {journal} {\bibinfo
  {journal} {Chaos: An interdisciplinary journal of nonlinear science}\
  }\textbf {\bibinfo {volume} {19}},\ \bibinfo {pages} {023117} (\bibinfo
  {year} {2009})}\BibitemShut {NoStop}%
\bibitem [{\citenamefont {Childs}\ and\ \citenamefont
  {Strogatz}(2008)}]{childs2008CIJNS}%
  \BibitemOpen
  \bibfield  {author} {\bibinfo {author} {\bibfnamefont {L.~M.}\ \bibnamefont
  {Childs}}\ and\ \bibinfo {author} {\bibfnamefont {S.~H.}\ \bibnamefont
  {Strogatz}},\ }\bibfield  {title} {\bibinfo {title} {Stability diagram for
  the forced {{Kuramoto}} model},\ }\href {https://doi.org/10.1063/1.3049136}
  {\bibfield  {journal} {\bibinfo  {journal} {Chaos: An Interdisciplinary
  Journal of Nonlinear Science}\ }\textbf {\bibinfo {volume} {18}},\ \bibinfo
  {pages} {043128} (\bibinfo {year} {2008})}\BibitemShut {NoStop}%
\bibitem [{\citenamefont {Laing}(2018)}]{laing2018JMN}%
  \BibitemOpen
  \bibfield  {author} {\bibinfo {author} {\bibfnamefont {C.~R.}\ \bibnamefont
  {Laing}},\ }\bibfield  {title} {\bibinfo {title} {The {{Dynamics}} of
  {{Networks}} of {{Identical Theta Neurons}}},\ }\bibfield  {journal}
  {\bibinfo  {journal} {The Journal of Mathematical Neuroscience}\ }\textbf
  {\bibinfo {volume} {8}},\ \href {https://doi.org/10.1186/s13408-018-0059-7}
  {10.1186/s13408-018-0059-7} (\bibinfo {year} {2018})\BibitemShut {NoStop}%
\bibitem [{\citenamefont {Goldobin}\ \emph {et~al.}(2018)\citenamefont
  {Goldobin}, \citenamefont {Tyulkina}, \citenamefont {Klimenko},\ and\
  \citenamefont {Pikovsky}}]{goldobin2018CIJNS}%
  \BibitemOpen
  \bibfield  {author} {\bibinfo {author} {\bibfnamefont {D.~S.}\ \bibnamefont
  {Goldobin}}, \bibinfo {author} {\bibfnamefont {I.~V.}\ \bibnamefont
  {Tyulkina}}, \bibinfo {author} {\bibfnamefont {L.~S.}\ \bibnamefont
  {Klimenko}},\ and\ \bibinfo {author} {\bibfnamefont {A.}~\bibnamefont
  {Pikovsky}},\ }\bibfield  {title} {\bibinfo {title} {Collective mode
  reductions for populations of coupled noisy oscillators},\ }\href
  {https://doi.org/10.1063/1.5053576} {\bibfield  {journal} {\bibinfo
  {journal} {Chaos: An Interdisciplinary Journal of Nonlinear Science}\
  }\textbf {\bibinfo {volume} {28}},\ \bibinfo {pages} {101101} (\bibinfo
  {year} {2018})}\BibitemShut {NoStop}%
\bibitem [{\citenamefont {Tyulkina}\ \emph {et~al.}(2018)\citenamefont
  {Tyulkina}, \citenamefont {Goldobin}, \citenamefont {Klimenko},\ and\
  \citenamefont {Pikovsky}}]{tyulkina2018PRL}%
  \BibitemOpen
  \bibfield  {author} {\bibinfo {author} {\bibfnamefont {I.~V.}\ \bibnamefont
  {Tyulkina}}, \bibinfo {author} {\bibfnamefont {D.~S.}\ \bibnamefont
  {Goldobin}}, \bibinfo {author} {\bibfnamefont {L.~S.}\ \bibnamefont
  {Klimenko}},\ and\ \bibinfo {author} {\bibfnamefont {A.}~\bibnamefont
  {Pikovsky}},\ }\bibfield  {title} {\bibinfo {title} {Dynamics of {{Noisy
  Oscillator Populations}} beyond the {{Ott-Antonsen Ansatz}}},\ }\bibfield
  {journal} {\bibinfo  {journal} {Physical Review Letters}\ }\textbf {\bibinfo
  {volume} {120}},\ \href {https://doi.org/10.1103/PhysRevLett.120.264101}
  {10.1103/PhysRevLett.120.264101} (\bibinfo {year} {2018})\BibitemShut
  {NoStop}%
\bibitem [{\citenamefont {Watanabe}\ and\ \citenamefont
  {Strogatz}(1993)}]{watanabe1993PRL}%
  \BibitemOpen
  \bibfield  {author} {\bibinfo {author} {\bibfnamefont {S.}~\bibnamefont
  {Watanabe}}\ and\ \bibinfo {author} {\bibfnamefont {S.~H.}\ \bibnamefont
  {Strogatz}},\ }\bibfield  {title} {\bibinfo {title} {Integrability of a
  globally coupled oscillator array},\ }\href
  {https://doi.org/10.1103/PhysRevLett.70.2391} {\bibfield  {journal} {\bibinfo
   {journal} {Physical Review Letters}\ }\textbf {\bibinfo {volume} {70}},\
  \bibinfo {pages} {2391} (\bibinfo {year} {1993})}\BibitemShut {NoStop}%
\bibitem [{\citenamefont {Goldobin}(2021{\natexlab{a}})}]{goldobin2021C}%
  \BibitemOpen
  \bibfield  {author} {\bibinfo {author} {\bibfnamefont {D.~S.}\ \bibnamefont
  {Goldobin}},\ }\bibfield  {title} {\bibinfo {title} {Mean-field models of
  populations of quadratic integrate-and-fire neurons with noise on the basis
  of the circular cumulant approach},\ }\href
  {https://doi.org/10.1063/5.0061575} {\bibfield  {journal} {\bibinfo
  {journal} {Chaos: An Interdisciplinary Journal of Nonlinear Science}\
  }\textbf {\bibinfo {volume} {31}},\ \bibinfo {pages} {083112} (\bibinfo
  {year} {2021}{\natexlab{a}})}\BibitemShut {NoStop}%
\bibitem [{\citenamefont {Binney}\ \emph {et~al.}(2001)\citenamefont {Binney},
  \citenamefont {Dowrick}, \citenamefont {Fisher},\ and\ \citenamefont
  {Newman}}]{binney2001}%
  \BibitemOpen
  \bibfield  {author} {\bibinfo {author} {\bibfnamefont {J.~J.}\ \bibnamefont
  {Binney}}, \bibinfo {author} {\bibfnamefont {N.~J.}\ \bibnamefont {Dowrick}},
  \bibinfo {author} {\bibfnamefont {A.~J.}\ \bibnamefont {Fisher}},\ and\
  \bibinfo {author} {\bibfnamefont {M.~E.~J.}\ \bibnamefont {Newman}},\
  }\href@noop {} {\emph {\bibinfo {title} {The {{Theory}} of {{Critical
  Phenomena}}: An Introduction to the Renormalization Group}}}\ (\bibinfo
  {publisher} {{Oxford Univ. Press}},\ \bibinfo {year} {2001})\BibitemShut
  {NoStop}%
\bibitem [{\citenamefont {Hohenberg}\ and\ \citenamefont
  {Halperin}(1977)}]{hohenberg1977RMP}%
  \BibitemOpen
  \bibfield  {author} {\bibinfo {author} {\bibfnamefont {P.~C.}\ \bibnamefont
  {Hohenberg}}\ and\ \bibinfo {author} {\bibfnamefont {B.~I.}\ \bibnamefont
  {Halperin}},\ }\bibfield  {title} {\bibinfo {title} {Theory of dynamic
  critical phenomena},\ }\href {https://doi.org/10.1103/RevModPhys.49.435}
  {\bibfield  {journal} {\bibinfo  {journal} {Reviews of Modern Physics}\
  }\textbf {\bibinfo {volume} {49}},\ \bibinfo {pages} {435} (\bibinfo {year}
  {1977})}\BibitemShut {NoStop}%
\bibitem [{\citenamefont {Ponce-Alvarez}\ \emph {et~al.}(2015)\citenamefont
  {Ponce-Alvarez}, \citenamefont {Deco}, \citenamefont {Hagmann}, \citenamefont
  {Romani}, \citenamefont {Mantini},\ and\ \citenamefont
  {Corbetta}}]{ponce-alvarez2015PCB}%
  \BibitemOpen
  \bibfield  {author} {\bibinfo {author} {\bibfnamefont {A.}~\bibnamefont
  {Ponce-Alvarez}}, \bibinfo {author} {\bibfnamefont {G.}~\bibnamefont {Deco}},
  \bibinfo {author} {\bibfnamefont {P.}~\bibnamefont {Hagmann}}, \bibinfo
  {author} {\bibfnamefont {G.~L.}\ \bibnamefont {Romani}}, \bibinfo {author}
  {\bibfnamefont {D.}~\bibnamefont {Mantini}},\ and\ \bibinfo {author}
  {\bibfnamefont {M.}~\bibnamefont {Corbetta}},\ }\bibfield  {title} {\bibinfo
  {title} {Resting-{{State Temporal Synchronization Networks Emerge}} from
  {{Connectivity Topology}} and {{Heterogeneity}}},\ }\href
  {https://doi.org/10.1371/journal.pcbi.1004100} {\bibfield  {journal}
  {\bibinfo  {journal} {PLOS Computational Biology}\ }\textbf {\bibinfo
  {volume} {11}},\ \bibinfo {pages} {e1004100} (\bibinfo {year}
  {2015})}\BibitemShut {NoStop}%
\bibitem [{\citenamefont {Cabral}\ \emph {et~al.}(2017)\citenamefont {Cabral},
  \citenamefont {Kringelbach},\ and\ \citenamefont {Deco}}]{cabral2017N}%
  \BibitemOpen
  \bibfield  {author} {\bibinfo {author} {\bibfnamefont {J.}~\bibnamefont
  {Cabral}}, \bibinfo {author} {\bibfnamefont {M.~L.}\ \bibnamefont
  {Kringelbach}},\ and\ \bibinfo {author} {\bibfnamefont {G.}~\bibnamefont
  {Deco}},\ }\bibfield  {title} {\bibinfo {title} {Functional connectivity
  dynamically evolves on multiple time-scales over a static structural
  connectome: {{Models}} and mechanisms},\ }\href
  {https://doi.org/10.1016/j.neuroimage.2017.03.045} {\bibfield  {journal}
  {\bibinfo  {journal} {NeuroImage}\ }\bibinfo {series} {Functional
  {{Architecture}} of the {{Brain}}},\ \textbf {\bibinfo {volume} {160}},\
  \bibinfo {pages} {84} (\bibinfo {year} {2017})}\BibitemShut {NoStop}%
\bibitem [{\citenamefont {Acebrón}\ \emph {et~al.}(2005)\citenamefont
  {Acebrón}, \citenamefont {Bonilla}, \citenamefont {Pérez~Vicente},
  \citenamefont {Ritort},\ and\ \citenamefont {Spigler}}]{acebron2005RMP}%
  \BibitemOpen
  \bibfield  {author} {\bibinfo {author} {\bibfnamefont {J.~A.}\ \bibnamefont
  {Acebrón}}, \bibinfo {author} {\bibfnamefont {L.~L.}\ \bibnamefont
  {Bonilla}}, \bibinfo {author} {\bibfnamefont {C.~J.}\ \bibnamefont
  {Pérez~Vicente}}, \bibinfo {author} {\bibfnamefont {F.}~\bibnamefont
  {Ritort}},\ and\ \bibinfo {author} {\bibfnamefont {R.}~\bibnamefont
  {Spigler}},\ }\bibfield  {title} {\bibinfo {title} {The {{Kuramoto}} model:
  {{A}} simple paradigm for synchronization phenomena},\ }\href
  {https://doi.org/10.1103/RevModPhys.77.137} {\bibfield  {journal} {\bibinfo
  {journal} {Reviews of Modern Physics}\ }\textbf {\bibinfo {volume} {77}},\
  \bibinfo {pages} {137} (\bibinfo {year} {2005})}\BibitemShut {NoStop}%
\bibitem [{\citenamefont {Dean}(1996)}]{dean1996JPAMG}%
  \BibitemOpen
  \bibfield  {author} {\bibinfo {author} {\bibfnamefont {D.~S.}\ \bibnamefont
  {Dean}},\ }\bibfield  {title} {\bibinfo {title} {Langevin equation for the
  density of a system of interacting {{Langevin}} processes},\ }\href
  {https://doi.org/10.1088/0305-4470/29/24/001} {\bibfield  {journal} {\bibinfo
   {journal} {Journal of Physics A: Mathematical and General}\ }\textbf
  {\bibinfo {volume} {29}},\ \bibinfo {pages} {L613} (\bibinfo {year}
  {1996})}\BibitemShut {NoStop}%
\bibitem [{Note1()}]{Note1}%
  \BibitemOpen
  \bibinfo {note} {The noise field $\eta (\phi , t)$ does depend on the angle
  value, while the individual noises $\eta _j(t)$ are just a function of time.
  Despite the similarity of notation, which reflects that the term $\protect
  \sqrt \rho \eta $ emerges from $\eta _j$, both terms are
  different.}\BibitemShut {Stop}%
\bibitem [{Note2()}]{Note2}%
  \BibitemOpen
  \bibinfo {note} {Can be directly obtained with the aid of computational
  algebra system. I used WolframEngine.}\BibitemShut {Stop}%
\bibitem [{\citenamefont {Hong}\ \emph {et~al.}(2015)\citenamefont {Hong},
  \citenamefont {Chaté}, \citenamefont {Tang},\ and\ \citenamefont
  {Park}}]{hong2015PRE}%
  \BibitemOpen
  \bibfield  {author} {\bibinfo {author} {\bibfnamefont {H.}~\bibnamefont
  {Hong}}, \bibinfo {author} {\bibfnamefont {H.}~\bibnamefont {Chaté}},
  \bibinfo {author} {\bibfnamefont {L.-H.}\ \bibnamefont {Tang}},\ and\
  \bibinfo {author} {\bibfnamefont {H.}~\bibnamefont {Park}},\ }\bibfield
  {title} {\bibinfo {title} {Finite-size scaling, dynamic fluctuations, and
  hyperscaling relation in the {{Kuramoto}} model},\ }\href@noop {} {\bibfield
  {journal} {\bibinfo  {journal} {Physical Review E}\ }\textbf {\bibinfo
  {volume} {92}} (\bibinfo {year} {2015})}\BibitemShut {NoStop}%
\bibitem [{\citenamefont {Buendía}\ \emph {et~al.}(2021)\citenamefont
  {Buendía}, \citenamefont {Villegas}, \citenamefont {Burioni},\ and\
  \citenamefont {Muñoz}}]{buendia2021PRR}%
  \BibitemOpen
  \bibfield  {author} {\bibinfo {author} {\bibfnamefont {V.}~\bibnamefont
  {Buendía}}, \bibinfo {author} {\bibfnamefont {P.}~\bibnamefont {Villegas}},
  \bibinfo {author} {\bibfnamefont {R.}~\bibnamefont {Burioni}},\ and\ \bibinfo
  {author} {\bibfnamefont {M.~A.}\ \bibnamefont {Muñoz}},\ }\bibfield  {title}
  {\bibinfo {title} {Hybrid-type synchronization transitions: {{Where}}
  incipient oscillations, scale-free avalanches, and bistability live
  together},\ }\href {https://doi.org/10.1103/PhysRevResearch.3.023224}
  {\bibfield  {journal} {\bibinfo  {journal} {Physical Review Research}\
  }\textbf {\bibinfo {volume} {3}},\ \bibinfo {pages} {023224} (\bibinfo {year}
  {2021})}\BibitemShut {NoStop}%
\bibitem [{\citenamefont
  {Goldobin}(2021{\natexlab{b}})}]{goldobin_mean-field_2021}%
  \BibitemOpen
  \bibfield  {author} {\bibinfo {author} {\bibfnamefont {D.~S.}\ \bibnamefont
  {Goldobin}},\ }\bibfield  {title} {\bibinfo {title} {Mean-field models of
  populations of quadratic integrate-and-fire neurons with noise on the basis
  of the circular cumulant approach},\ }\href
  {https://doi.org/10.1063/5.0061575} {\bibfield  {journal} {\bibinfo
  {journal} {Chaos: An Interdisciplinary Journal of Nonlinear Science}\
  }\textbf {\bibinfo {volume} {31}},\ \bibinfo {pages} {083112} (\bibinfo
  {year} {2021}{\natexlab{b}})}\BibitemShut {NoStop}%
\bibitem [{\citenamefont {Goldobin}\ \emph {et~al.}(2021)\citenamefont
  {Goldobin}, \citenamefont {di~Volo},\ and\ \citenamefont
  {Torcini}}]{goldobin_reduction_2021}%
  \BibitemOpen
  \bibfield  {author} {\bibinfo {author} {\bibfnamefont {D.~S.}\ \bibnamefont
  {Goldobin}}, \bibinfo {author} {\bibfnamefont {M.}~\bibnamefont {di~Volo}},\
  and\ \bibinfo {author} {\bibfnamefont {A.}~\bibnamefont {Torcini}},\
  }\bibfield  {title} {\bibinfo {title} {Reduction {Methodology} for
  {Fluctuation} {Driven} {Population} {Dynamics}},\ }\href
  {https://doi.org/10.1103/PhysRevLett.127.038301} {\bibfield  {journal}
  {\bibinfo  {journal} {Physical Review Letters}\ }\textbf {\bibinfo {volume}
  {127}},\ \bibinfo {pages} {038301} (\bibinfo {year} {2021})},\ \bibinfo
  {note} {publisher: American Physical Society}\BibitemShut {NoStop}%
\bibitem [{\citenamefont {Dolmatova}\ \emph {et~al.}(2023)\citenamefont
  {Dolmatova}, \citenamefont {Tyulkina},\ and\ \citenamefont
  {Goldobin}}]{dolmatova_circular_2023}%
  \BibitemOpen
  \bibfield  {author} {\bibinfo {author} {\bibfnamefont {A.~V.}\ \bibnamefont
  {Dolmatova}}, \bibinfo {author} {\bibfnamefont {I.~V.}\ \bibnamefont
  {Tyulkina}},\ and\ \bibinfo {author} {\bibfnamefont {D.~S.}\ \bibnamefont
  {Goldobin}},\ }\bibfield  {title} {\bibinfo {title} {Circular cumulant
  reductions for macroscopic dynamics of oscillator populations with
  non-{Gaussian} noise},\ }\href {https://doi.org/10.1063/5.0159982} {\bibfield
   {journal} {\bibinfo  {journal} {Chaos: An Interdisciplinary Journal of
  Nonlinear Science}\ }\textbf {\bibinfo {volume} {33}},\ \bibinfo {pages}
  {113102} (\bibinfo {year} {2023})}\BibitemShut {NoStop}%
\end{thebibliography}
\end{document}


\title{Supplementary information for\\ A mesoscopic theory for stochastic coupled oscillators}

\author{Victor Buendía}
 \email{vbuendiar@onsager.ugr.es}
\affiliation{Department of Computer Science, University of Tübingen, Tübingen, Germany}
\affiliation{Max Planck Institute for Biological Cybernetics, Tübingen, Germany}
\affiliation{Department of Computing Sciences, Bocconi University, Milan, Italy}

\date{\today}

\maketitle
\appendix


\section{Dean-Kawasaki formalism}

For completeness, I offer here an informal derivation of the Dean-Kawasaki equation  for oscillators. A more formal discussion on the procedure can be found in the original paper by Dean \cite{dean1996JPAMG}.

Consider an oscillator model given by a Langevin equation in the Itô sense,

\begin{equation}
    \dot \phi_i = F_i[\{\phi_j\}_{j=1,2,\ldots,N}] + \sigma \xi_i(t).
\end{equation}

The goal is to obtain the dynamics of the density $\rho(\phi, t)$ of oscillators in the interval $[\phi, \phi+d\phi]$. The first step is to define an individual density, $\rho_i (\phi, t) = \delta(\phi_i - \phi)$, where $\delta(\phi)$ represents the Dirac delta distribution. The total density of the system can thus be obtained just as $\rho(\phi, t) = N^{-1} \sum_i \rho_i(\phi, t)$, given that there is a sufficient amount of oscillators ($N \gg 1$). Changing variable from $\phi_i$ to $\rho_i$ (applying Itô's lemma) one has

\begin{equation}
    \partial_t \rho_i = \partial_{\phi_i} \rho_i \dot \phi_i + \frac{\sigma^2}{2} \partial_{\phi_i} ^2 \rho_i. 
\end{equation}

Notice that $\partial_{\phi_i} \rho_i(\phi, t) = -\partial_\phi \rho_i (\phi, t)$, due to the properties of the Dirac delta. Moreover, terms $\phi_i$ in the equations can be changed to $\phi$, since they are fixed by the delta, 

\begin{equation}
    \partial_t \rho_i = \partial_\phi \left( \rho_i F_i(\phi, \{\phi_j\}) \right) +  \sigma \partial_\phi \left( \rho_i \xi_i \right) + \frac{\sigma^2}{2} \partial_\phi ^2 \rho_i. 
\end{equation}

Then, one has to sum over all indices $i$ and multiply by $N^{-1}$, which leads to   

\begin{equation}
    \partial_t \rho = \partial_\phi \left( \frac{1}{N}\sum_i \rho_i F_i(\phi, \{\phi_j\}) \right) + \frac{\sigma^2}{2} \partial_\phi ^2 \rho + \frac{\sigma}{N}\partial_\phi \left( \sum_i \rho_i \xi_i \right). 
\end{equation}

If the system size is large enough, the noise term can be approximated by $N^{-1/2} \partial_\phi \left( \eta(\phi, t) \sqrt{\rho(\phi, t)} \right)$. It is possible to show that the correlations of such noise are identical to the correlations of $N^{-1} \partial_\phi \left( \sum_i \rho_i \xi_i \right)$, as done by Dean \cite{dean1996JPAMG}. Notice the extra $1/\sqrt{N}$ factor, which comes from the fact that we are normalizing the density. 

The simplification of the deterministic part depends on the considered system; in the case of mean-field coupled oscillators, interaction terms such as $\sum_j (\phi_j - \phi_i)$  can be rewritten using the Kuramoto-Daido parameters, $Z_k$. Thus, a large family of oscillators can be rewritten into the form $F_i(\phi, \{\phi_j\}) \equiv F(\phi, \{Z_k\})$ and the resulting equation can be simplified into  

\begin{equation}
    \partial_t \rho = \partial_\phi \left( \rho F(\phi, \{Z_k \}) \right) + \frac{\sigma^2}{2} \partial_\phi ^2 \rho +  \frac{\sigma}{\sqrt{N}} \partial_\phi \left( \eta(\phi, t) \sqrt{\rho} \right). 
    \label{eq:dk}
\end{equation}

which is the Dean-Kawasaki equation for oscillators considered in the main text.


\section{Fluctuations of the Kuramoto-Daido parameters}

In the continuous limit, the Kuramoto-Daido parameters are defined by 

\begin{equation}
    Z_k = \int _0 ^{2\pi} d\phi \rho(\phi, t) e^{ik\phi}.
    \label{eq:kuramoto-daido}
\end{equation}

The real and imaginary parts of $Z_k$ will be denoted by $Z_k(t) = x_k(t) + i y_k(t)$. 

The temporal evolution $\dot Z_k$ can be obtained deriving with respect to time and inserting eq.~(\ref{eq:dk}). The result of the deterministic part is identical to the procedure obtained using a Fokker-Planck equation. Here, I work out the stochastic term,

\begin{equation}
    \xi_k(t) = \frac{\sigma}{\sqrt{N}} \int _0 ^{2\pi} d\phi \partial_\phi \left( \eta(\phi, t) \sqrt{\rho} \right) e^{ik\phi}.
\end{equation}

The main idea is to integrate by parts, taking $u=e^{ik\phi}$ and $dv=d\phi \partial_\phi \left( \eta(\phi, t) \sqrt{\rho} \right)$. In general, the application of integration by parts in a stochastic Itô integral implies considering the covariance between the two processes\cite{gardiner_stochastic_2009}. In this case, $\rho$ and $\xi$ are uncorrelated and we can proceed as always,

\begin{equation}
    \xi_k(t) = \frac{\sigma}{\sqrt{N}}\left . e^{ik\phi}\eta(\phi, t) \sqrt{\rho} \right|_{0} ^{2\pi} - \frac{ik\sigma}{\sqrt{N}} \int _0 ^{2\pi}  d\phi  \left( \eta(\phi, t) \sqrt{\rho} \right) e^{ik\phi}
\end{equation}

The first term vanishes due to the periodicity on $\phi$. The result is a noise term which is given by the integration over all the $\phi$-space, 

\begin{equation}
    \xi_k(t) = -\frac{i k \sigma}{\sqrt{N}} \int _0 ^{2\pi}  d\phi   \eta(\phi, t) \sqrt{\rho}  e^{ik\phi}. 
\end{equation}

This noise is a complex number with real and imaginary parts $\xi_k = \xi_{x,k}(t) + i \xi_{y,k}(t)$. We need to compute the covariance matrix between all the elements to fully characterize it. 

The first element is $\left\langle \xi_{j,x}\left(t\right)\xi_{k,x}\left(t'\right)\right\rangle$. This can be done by direct substitution, 

\begin{align}
    \left\langle \xi_{j,x}\left(t\right)\xi_{k,x}\left(t'\right)\right\rangle &=\frac{kj\sigma^{2}}{N}\int d\phi d\phi'\sin\left(j\phi\right)\sin\left(k\phi'\right)\sqrt{\rho\left(\phi,t\right)\rho\left(\phi',t'\right)}\left\langle \eta\left(\phi,t\right)\eta\left(\phi',t'\right)\right\rangle = \nonumber \\
    &=\frac{kj\sigma^{2}}{N}\delta\left(t-t'\right)\int d\phi\sin\left(j\phi\right)\sin\left(k\phi\right)\rho\left(\phi,t\right)
    =\frac{kj\sigma^{2}}{N}\delta\left(t-t'\right)\cdot\frac{1}{2}\left(x_{j-k}-x_{j+k}\right).
    \label{eq:ex_computation_corr}
\end{align}

In the first step, the correlation between the two noise fields yields a $\delta(\phi'-\phi)\delta(t'-t)$ which is used to eliminate the integral in $\phi'$, simplifying the square root with the density terms. Then, $\sin A\sin B$ is written as $(\cos(A-B) - \cos(A+B))/2$, which allows us to identify the remaining integral with $x_k(t)$, the real part of the Kuramoto-Daido parameter $Z_k(t)$.

One can proceed analogously with all the other terms. The results are those written in the main text,

\begin{subequations}
\begin{align}
    \left\langle \xi_{j,x}(t)\xi_{k,x}(t')\right\rangle  &= \frac{kj\sigma^{2}}{2N}\delta\left(t-t'\right)\left(x_{j-k}-x_{j+k}\right), \\
    \left\langle \xi_{j,y}(t)\xi_{k,x}(t')\right\rangle  &= \frac{kj\sigma^{2}}{2N}\delta\left(t-t'\right)\left(y_{j-k}-y_{j+k}\right), \\
    \left\langle \xi_{j,y}(t)\xi_{k,y}(t')\right\rangle  &= \frac{kj\sigma^{2}}{2N}\delta\left(t-t'\right)\left(x_{j-k}+x_{j+k}\right).
    \label{corrs}
\end{align}
\end{subequations}

\subsection{Extension to multiplicative noise}

It is possible to generalize our setup to work with multiplicative noise, 

\begin{equation}
    \dot \phi_i = F_i[\{\phi_j\}_{j=1,2,\ldots,N}] + G(\{\phi_j\}_{j=1,2,\ldots,N}) \xi_i(t).
\end{equation}

just by writing a generalized Dean-Kawasaki equation. This can be done in a naive heuristic way, just looking at how the diffusion term changes in the Fokker-Planck, and doing the same for the Dean-Kawasaki, leading to 

\begin{equation}
    \partial_t \rho = \partial_\phi \left[ \rho F(\phi, \{Z_k \}) \right] + \frac{1}{2} \left[ \partial_\phi ^2 G(\phi, \{ Z_k \}) \rho \right]+  \frac{1}{\sqrt{N}} \partial_\phi \left( G(\phi, \{ Z_k \}) \eta(\phi, t) \sqrt{\rho} \right). 
\end{equation}

One can compute the correlations of the effective noise term and see that they coincide with those of the microscopic process, as it happened in the additive case. A formal derivation of the generalized Dean-Kawasaki derivation has also been done by Djurdjevac \emph{et al.} \cite{djurdjevac_conrad_feedback_2022}. 

Observe that $G$ should be written as a function of the angle itself and the mean-field Kuramoto-Daido parameters for the method to work. Additionally, we will ask $G(\phi, \{Z_k\})$ to be periodic in $\phi$, so the boundary terms vanish when the integration by parts is performed. This option is the most natural for oscillators, and it could be relaxed by carrying the boundary terms.

Then our noise can be generalized to 

\begin{equation}
    \xi_k(t) = -\frac{i k \sigma}{\sqrt{N}} \int _0 ^{2\pi}  d\phi G(\phi, \{Z_k\})  \eta(\phi, t) \sqrt{\rho}  e^{ik\phi}. 
\end{equation}

and solving the integral depends on the exact shape of $G$. Since we asked for periodicity in $\phi$, it can be thus decomposed in a Fourier expansion; so the integral will always include the product of three trigonometric functions, that can be re-arranged as sum and difference of sines and cosines. These, together with the density, can be identified as the Kuramoto-Daido parameters, as in the additive noise case.

As a pedagogical example, one could consider $G(\phi, \{Z_k\}) = R_1 \sin(\phi - \psi_1)$, so individual fluctuations vanish for oscillators which are close to the mean phase, as well as for asynchronous systems. Then,

\begin{align}
    R_1 \sin(j\phi)\sin(k\phi)\sin(\phi - \psi_1) =& \sin(j\phi)\sin(k\phi) \left[ x_1 \sin\phi - y_1 \cos\phi\right] = \nonumber \\
    =& x_1 \left[ \sin\left((1+j-k)\phi \right) + \sin\left((1-j+k)\phi \right) - \sin\left((1-j-k)\phi \right) - \sin\left((1+j+k)\phi \right) \right] - \nonumber \\
    -& y_1 \left[ \cos\left((1+j-k)\phi \right) + \cos\left((1-j+k)\phi \right) - \cos\left((1-j-k)\phi \right) - \cos\left((1+j+k)\phi \right) \right]. 
\end{align}

Hence, one can identify 

\begin{align}
    \left\langle \xi_{j,x}(t)\xi_{k,x}(t')\right\rangle  &= \frac{kj\sigma^{2}}{2N}\delta\left(t-t'\right) \left[x_1\left(y_{1+j-k}+y_{1-j+k}-y_{1+k+j}-y_{1-k-j}\right)-y_1\left(x_{1+j-k}+x_{1-j+k}-x_{1+k+j}-x_{1-k-j}\right)  \right]
\end{align}

where fluctuations vanish in both the asynchronous and the perfectly synchronous state, as one would have expected from the microscopic shape of the fluctuations.

\subsection{Extension to randomly distributed frequencies}

It is possible to extend our results to populations of oscillators whose frequencies are distributed with a unimodal probability distribution $g(\omega)$, as in the classical Kuramoto model. In order to do that, we follow the same procedure as before: write the Dean-Kawasaki equation, expand the density in Fourier series, and write dynamical equations for the Fourier coefficients. 

For the first step, the construction of the Dean-Kawasaki equation is almost identical, except that the densities now depend on $\omega$ through the microscopic densities $\rho_i(\omega, \phi, t)$. Then, the effective noise has to depend on the frequency, promoting the noise field $\eta(\phi, t)$ to $\eta(\omega, \phi, t)$.  If all oscillator frequencies were sampled independently, then 

\begin{equation}
\langle \eta(\omega, \phi, t) \eta(\omega', \phi', t') \rangle = \delta(\omega - \omega') \delta(\phi - \phi') \delta(t-t'). 
\end{equation}

In the homogeneous case, the Kuramoto-Daido parameters directly coincided with the Fourier coefficients.  When the heterogeneity in frequencies is considered, the Fourier coefficients $\alpha_k(t, \omega)$ depend also on the frequency. Hence, they cannot coincide with the order parameters. They are, however, related by the integral

\begin{equation}
    Z_k(t) = \int_{-\infty} ^{+\infty} d\omega g(\omega) \alpha_k(\omega,t).
    \label{eq:aktozk}
\end{equation}

as shown originally by Ott and Antonsen\cite{ott2008C, ott2009CIJNS}. The way to generalize the theory is to realize that the computations above are equally valid, just substituting $Z_k(t)$ by $\alpha_k(\omega, t)$.  since by definition

\begin{equation}
    \alpha_k(\omega, t) = \int _0 ^{2\pi} \rho(\omega, \phi, t) e^{ik\phi} d\phi, 
\end{equation}

so it suffices to derive with respect to time and integrate by parts, analogously to what we did before. The definition of the noise does not change, but the correlations will yield terms $\alpha_k$ instead, since e.g. eq.(\ref{eq:ex_computation_corr}) gives  

\begin{align}
    \left\langle \xi_{j,x}\left(\omega, t\right)\xi_{k,x}\left(\omega', t'\right)\right\rangle 
    &=\frac{kj\sigma^{2}}{N}\delta\left(t-t'\right)\delta(\omega-\omega') \int d\phi\sin\left(j\phi\right)\sin\left(k\phi\right)\rho\left(\omega, \phi,t\right) = \nonumber \\
    &=\frac{kj\sigma^{2}}{N}\delta\left(t-t'\right)\delta(\omega-\omega') \cdot\frac{1}{2}\text{Re}\left(\alpha_{j-k}-\alpha_{j+k}\right).
\end{align}

Finally, there is an extra step to relate $\alpha_k$ to $Z_k$, by solving the integral over $g(\omega)$. This integral cannot be explicitly solved always. In practice, one may need to integrate the Langevin equations self-consistently. In simple cases, such as the Lorentzian distribution, eq.(\ref{eq:aktozk}) can be integrated using residues \cite{ott2008C}. For example, for a Lorentzian with mean zero and unit dispersion,

\begin{equation}
    g(\omega) = \frac{1}{\pi(\omega ^2+1)}
\end{equation}

one gets $Z_k(t) = \bar \alpha(-ik, t)$, with the bar meaning complex conjugate. This substitution allows one to work directly with the Kuramoto-Daido parameters instead of the Fourier coefficients. Another simple case, $g(\omega)=\delta(\omega-\omega_0)$ immediately recovers our previous results.


\section{Change of variables to amplitude fluctuations}

When studying the synchronization of oscillators, one is usually interested in the value of the amplitude of $Z_k = R_k e^{i\psi_k}$, which indicates if the system is fully asynchronous, $R_k = 0$, or synchronous, $R_k = 1$. In many cases, the Kuramoto order parameter $R_1$ is sufficient to qualitatively describe the behaviour of the system. 

Thus, the next step is to change variables in the equations for $x_k, y_k$ to obtain the time evolution of $R_k = \sqrt{x_k ^2 + y_k ^2}$. This is not a trivial task, since the noise is multiplicative and highly correlated. I include here the detailed procedure for the change of variables, finding the deterministic drift emerging from the application of Itô's lemma, and the shape of the correlation in the new variables. In general, the change of variables formula can be written as 

\begin{equation}
    dR_{k}=\sum_{k'}\left(\frac{\partial R_{k}}{\partial x_{k'}}dx_{k'}+\frac{\partial R_{k}}{\partial y_{k'}}dy_{k'}\right)+\underbrace{\frac{1}{2}\sum_{ij}\left[\frac{\partial^{2}R_{k}}{\partial x_{i}\partial x_{j}}dx_{i}dx_{j}+2\frac{\partial^{2}R_{k}}{\partial x_{i}\partial y_{j}}dx_{i}dy_{j}+\frac{\partial^{2}R_{k}}{\partial y_{i}\partial y_{j}}dy_{i}dy_{j}\right]}_{\Lambda_{R,k}/N},
    \label{eq:itolemma}
\end{equation}

where the differentials can be identified with the elements of the correlation matrix at order $dt$, i.e., $dx_i dx_j \sim dt \langle \xi_{i,x}(t) \xi_{j,x}(t)\langle$ and so on. The deterministic drift to the equations comes from the second derivatives. Let me start analysing how the noise term changes. This comes from the first term in the equation, after expanding $dx_{k'}, dy_{k'}$. Thus, fluctuations in amplitude will be of the form $\xi_{k,R}=\cos\psi_{k}\xi_{k,x}+\sin\psi_{k}\xi_{k,y}$. This allows us to compute the correlation matrix in polar coordinates,

\begin{align}
\left\langle \xi_{k,R}\left(t\right)\xi_{k',R}\left(t'\right)\right\rangle &= \cos\psi_{k}\cos\psi_{k'}\left\langle \xi_{x}\xi_{x}'\right\rangle +\cos\psi_{k}\sin\psi_{k'}\left\langle \xi_{x}\xi_{y}'\right\rangle +\sin\psi_{k}\cos\psi_{k'}\left\langle \xi_{y}\xi_{x}'\right\rangle +\sin\psi_{k}\sin\psi_{k'}\left\langle \xi_{y}\xi_{y}'\right\rangle = \nonumber \\
(i)& = \frac{kk'\sigma^{2}}{2N}\delta\left(t-t'\right) \left[ \cos\psi_{k}\cos\psi_{k'}\left(x_{k'-k}-x_{k'+k}\right)+\cos\psi_{k}\sin\psi_{k'}\left(y_{k'-k}-y_{k'+k}\right)+\right. \nonumber \\
&+ \left. \sin\psi_{k}\cos\psi_{k'}\left(y_{k-k'}-y_{k+k'}\right)+\sin\psi_{k}\sin\psi_{k'}\left(x_{k'-k}+x_{k'+k}\right)\right] \nonumber = \\
(ii)&=C_{kk'} \left[ x_{k'-k}\left(\cos\psi_{k}\cos\psi_{k'}+\sin\psi_{k}\sin\psi_{k'}\right)+x_{k'+k}\left(\sin\psi_{k}\sin\psi_{k'}-\cos\psi_{k}\cos\psi_{k'}\right) \right. \nonumber \\
&+ \left.  \cos\psi_{k}\sin\psi_{k'}\left(y_{k'-k}-y_{k'+k}\right)+\sin\psi_{k}\cos\psi_{k'}\left(-y_{k'-k}-y_{k+k'}\right) \right] = \nonumber \\
(iii)&= C_{kk'} \left[ x_{k'-k}\cos\left(\psi_{k}-\psi_{k'}\right)-x_{k'+k}\cos\left(\psi_{k'}+\psi_{k}\right)+ y_{k'-k}\sin\left(\psi_{k'}-\psi_{k}\right)-y_{k'+k}\sin\left(\psi_{k'}+\psi_{k}\right) \right]= \nonumber \\
(iv)&=\frac{kk'\sigma^{2}}{2N}\delta\left(t-t'\right)\left[R_{k-k'}\cos\left(\psi_{k'-k}-\left(\psi_{k'}-\psi_{k}\right)\right)-R_{k+k'}\cos\left(\psi_{k'+k}-\left(\psi_{k'}+\psi_{k}\right)\right)\right].
\end{align}

where I have (i) substituted the elements of the correlation matrix, (ii) re-arranged them, (iii) used that $\bar Z_k = Z_{-k}$ and employed trigonometrical identities, and finally (iv) substituted $x_k = R_k \cos\psi_k, y_k = R_k \sin\psi_k$ and further simplified the trigonometrical functions. An analogous procedure can be followed in order to determine that 

\begin{subequations}
\begin{align}
    \left\langle \xi_{k,R}\left(t\right)\xi_{k'\psi}\left(t'\right)\right\rangle &= \frac{kk'\sigma^{2}}{2NR_k}\delta\left(t-t'\right) \left[R_{k'-k}\sin\left(\psi_{k'-k}+\psi_{k}-\psi_{k'}\right)-R_{k'+k}\sin\left(\psi_{k'+k}-\left(\psi_{k}+\psi_{k'}\right)\right)\right], \\
    \left\langle \xi_{k,\psi}\left(t\right)\xi_{k',\psi}\left(t'\right)\right\rangle &= \frac{kk'\sigma^{2}}{2NR_{k}R_{k'}}\delta\left(t-t'\right) \left[R_{k'-k}\cos\left(\psi_{k'-k}-\left(\psi_{k'}-\psi_{k}\right)\right)+R_{k'+k}\cos\left(\psi_{k'+k}-\left(\psi_{k'}+\psi_{k}\right)\right)\right],
\end{align}
\end{subequations}

for the correlations with the angular variable. Finally, let me work out the deterministic drifts coming from the Itô's lemma. These arise from the the second derivatives of eq.~(\ref{eq:itolemma}), which can be further simplified,

\begin{subequations}
\begin{align}
     \frac{\partial^{2}R_{k}}{\partial x_{i}\partial x_{j}}=&\delta_{ik}\delta_{jk}\frac{\partial^{2}R_{k}}{\partial x_{k}^{2}} = \delta_{ik}\delta_{jk}\frac{\sin^{2}\psi_{k}}{R_{k}}, \\
    \frac{\partial^{2}R_{k}}{\partial y_{i}\partial y_{j}}=&\delta_{ik}\delta_{jk}\frac{\cos^{2}\psi_{k}}{R_{k}}, \\
     \frac{\partial^{2}R_{k}}{\partial x_{i}\partial y_{j}}=&-\delta_{ik}\delta_{jk}\frac{\sin\psi_{k}\cos\psi_{k}}{R_{k}}.
\end{align}
\end{subequations}

Introducing these expressions for the second derivatives into eq.~(\ref{eq:itolemma}), one gets

\begin{align}
    &\frac{\Lambda_{R,k}}{N} = \frac{1}{2}\sum_{ij}\left[\frac{\partial^{2}R_{k}}{\partial x_{i}\partial x_{j}}dx_{i}dx_{j}+2\frac{\partial^{2}R_{k}}{\partial x_{i}\partial y_{j}}dx_{i}dy_{j}+\frac{\partial^{2}R_{k}}{\partial y_{i}\partial y_{j}}dy_{i}dy_{j}\right]= \nonumber \\
    (i)&= dt\frac{\sin^{2}\psi_{k}}{2R_{k}}\left(\frac{k^{2}\sigma^{2}}{2N}\left(1-x_{2k}\right)\right)-dt\frac{\sin\psi_{k}\cos\psi_{k}}{R_{k}}\left(\frac{k^{2}\sigma^{2}}{2N}\left(0-y_{2k}\right)\right)+dt\frac{\cos^{2}\psi_{k}}{2R_{k}}\left(\frac{k^{2}\sigma^{2}}{2N}\left(1+x_{2k}\right)\right)dt= \nonumber \\
    (ii)&= \frac{k^{2}\sigma^{2}}{4NR_{k}}\left(1+x_{2k}\left(\cos^{2}\psi_{k}-\sin^{2}\psi_{k}\right)+2\sin\psi_{k}\cos\psi_{k}y_{2k}\right)dt = \nonumber \\
    (iii)&=\frac{k^{2}\sigma^{2}}{4NR_{k}}\left(1+R_{2k}\cos\left(\psi_{2k}-2\psi_{k}\right)\right)dt.
\end{align}

where in this case I have (i) substituted the second derivatives and the values of the correlation matrix, (ii) simplified the expression and (iii) reduced the trigonometrical relations. It is possible to repeat this procedure for the angle variable, which leads to a deterministic drift 

\begin{equation}
    \frac{\Lambda_{\psi, k}}{N} = \frac{1}{2}\sum_{ij}\left[\frac{\partial^{2}\psi_{k}}{\partial x_{i}\partial x_{j}}dx_{i}dx_{j}+2\frac{\partial^{2}\psi_{k}}{\partial x_{i}\partial y_{j}}dx_{i}dy_{j}+\frac{\partial^{2}\psi_{k}}{\partial y_{i}\partial y_{j}}dy_{i}dy_{j}\right]=
    \frac{R_{2k}k^{2}\sigma^{2}}{2NR_{k}^{2}}\left[\sin\left(\psi_{2k}-2\psi_{k}\right)\right]dt.
\end{equation}

Summarizing, we find that the deterministic drift terms scale with $\mathcal(1/N)$, making them negligible in the thermodynamic limit. Observe that $\Lambda_{R,k}$ contains a $1/R_k$ term which prevents the Kuramoto-Daido parameters from vanishing for a finite-size, causing the residual synchronization found in simulations.


\section{Closed-form solutions for the stochastic Kuramoto model}

In this section, I report closed formulas for the Kuramoto order parameter, as obtained from the theory. 

\subsection{Full amplitude recurrence in the thermodynamic limit}

The first case which can be analytically solved is the average of the Kuramoto-Daido parameters in the thermodynamic limit, i.e., when the deterministic drift vanishes. When one looks for the stationary solutions of $dZ_k/dt$, it is faced with the recurrence 

\begin{equation}
    \frac{J}{2}kR_1 \left(R_{k-1}-R_{k+1}\right) -\frac{\sigma^2 k^2}{2}R_k = 0.
\end{equation}

This is a generalization of the Ott-Antonsen ansatz for the amplitudes, that can be written also as a recurrence $R_{k+1} - R_k R_1 = 0$. 

The general solution of the recurrence can be written as a function of modified Bessel functions,

\begin{align}
      R_{k} = \tilde R \left[ K_{k}\left(-\tilde R_1 \right)\left(I_{1}\left(\tilde R\right)-R_1 I_{0}\left(\tilde R\right)\right) - I_{k}\left(\tilde R \right)\left(K_{1}\left(-\tilde R \right)-R K_{0}\left(-\tilde R\right)\right)\right]
      \label{eq:rec-solution}
\end{align}

where $\tilde R=2JR_1/\sigma^2$, and $I_{k}\left(z\right), K_{k}\left(z\right)$, the modified Bessel functions of first and second type, respectively. Observe that $R_1$ is not specified, since one needs two values to start the recurrence.  By definition, one has $R_0 = 1$, but $R_1$ is arbitrary. A way to obtain the Kuramoto order parameter $R_1$ is to impose a closure, i.e., to set $R_{k>k^*}=0$. This extra condition will fix $R_1$.

When one chooses $R_6 = 0$, it leads to 

\begin{equation}
    R_1=\frac{\sqrt{2}\sigma^{2}}{J}\sqrt{\frac{24\sigma^{2}-9J-\sqrt{51J^{2}-132J\sigma^{2}+306\sigma^{4}}}{J-9\sigma^{2}}},
\end{equation}

which is the expression shown in the main text. Tie formula is not exact, since it is just a closure up to the sixth harmonic. However, it allows us to fit the the values of the Kuramoto order parameter in a way which is even more precise than taking closing for two circular cumulants, even for relatively high values of $\sigma$. Figure \ref{fig:rfs_vs_data}a shows the same graph shown in the main text, but for a larger value of $\sigma$, demonstrating the accuracy of the formula.

As discussed in the main text, if better values of $R_1$ are desired, one just needs to numerically find the solution of $R_{k^*} = 0$ for the desired harmonic $k^*$ using eq.~(\ref{eq:rec-solution}), which is easier than solving the differential equations. Moreover, using asymptotic expansions of the modified Bessel functions for large $k$, in principle it should be possible to obtain closed-form approximations for $R_1$, applying a closure for large values of $k^*$ effectively.

\subsection{Ott-Antnsonsen solution for finite sizes}

If one assumes the Ott-Antonsen ansatz ($Z_k = Z_1 ^k$) in the case of the Kuramoto model, the stochastic equations read

\begin{subequations}
\begin{align}
    \dot{R}=&	\frac{1}{2}\left(J-\sigma^{2}\right)R-\frac{J}{2}R^{3}+\frac{\sigma^{2}}{2NR}\left(1+R^{2}\right)+\sigma\sqrt{\frac{1-R^{2}}{2N}}\xi_{R}, \\
\dot{\psi}=&	\omega+\sigma\sqrt{\frac{1}{2N}\left(1+\frac{1}{R^{2}}\right)}\xi_{\psi}\left(t\right).
\end{align}
\end{subequations}

where the indices have been dropped for the ease of reading,  $R_1 \equiv R, \psi_1 \equiv \psi$.  This allows us to perform an analisis of the finite-size, stochastic, Kuramoto model. For example, taking averages and assuming that $ \left\langle R^{2}\right\rangle =\left\langle R\right\rangle ^{2}$, one finds that the stationary state for the amplitude happens at 

\begin{equation}
    R^*(N)=\sqrt{\frac{JN-\left(N-1\right)\sigma^{2}+\sqrt{4JN\sigma^{2}+\left[JN-\left(N-1\right)\sigma^{2}\right]^{2}}}{2JN}}.
    \label{eq:rfinitesize}
\end{equation}

\begin{figure*}[!htpb]
\includegraphics[width=\columnwidth]{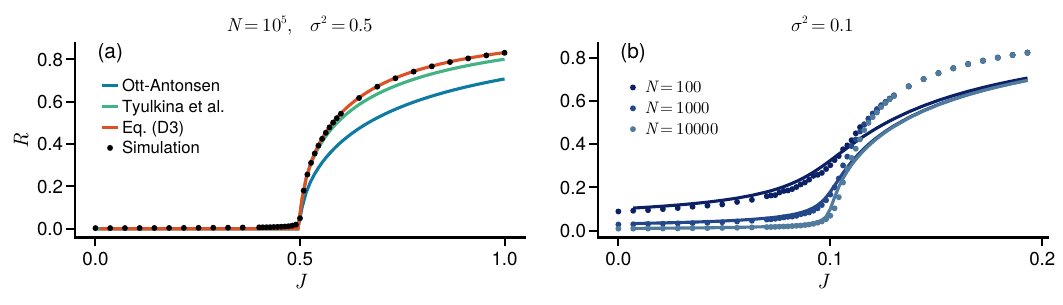}
\caption{ \textbf{Exactness of the closed-form solutions.} (a) The same simulations shown in the main text, but for $\sigma^2 = 0.5$. These show that the theory captures well the phenomenology also for larger noise intensities. (b) Analytical stationary solutions of eq.(\ref{eq:rfinitesize}) capture the finite size residual synchronization in the incoherent state.} \label{fig:rfs_vs_data}
\end{figure*}

Letting $N\to+\infty$ one recovers the classical thermodynamic limit, $\lim _{N\to+\infty} R^{*}(N) = \theta(J-\sigma^2) \sqrt{1-\frac{\sigma^{2}}{J}}$, being $\theta(x)$ the Heaviside step function. Eq. (\ref{eq:rfinitesize}) is plotted against simulations of the Kuramoto model for different sizes in Fig. \ref{fig:rfs_vs_data}. One can see that in the asynchronous phase, the equation captures very well the behaviour of the full model, failing in the synchronous one. Since we  are using the Ott-Antonsen ansatz, it is expected for the model to fail in the synchronous regime; however, the finite size drift term in conjunction with the ansatz is all needed to capture the residual synchronization seen in finite-size systems at low couplings.

\section{Simulation protocol}

The simulation of the mesosocopic Langevin equations poses several practical difficulties, which need some extra effort to be addressed. For this section, I will focus on the amplitude equations, although the same considerations need to be done, e.g., for the full system in cartesian coordinates. 

Formally, there are infinite equations, so the first step is to decide the number $k^*$ of harmonics to integrate and how to close the system. In simulations, one usually sets $Z_{k>k^*}=0$. Thus, the covariance matrix $\hat M$ between elements $\langle \xi_{R,k}(t) \xi_{R, k'}(t) \rangle$ will be of size $k^* \times k^*$, and $k^*$ random numbers must be generated. Observe that the covariance matrix depends on all $R_k$, and therefore it must be recomputed at each timestep. 

Once the covariance matrix has been updated, it can be used to obtain correlated Gaussian random numbers. The classical algorithm to do so is to perform a Cholesky decomposition $\hat M = \hat g \hat g^\dagger$, and multiply the decomposed matrix $\hat g$ by a vector of Gaussian variables $\vec \xi$ with mean zero and unit variance.

The main problem is that our covariance matrix might be non-positive definite, due to the finite size of the integration step and integration error. Intuitively speaking, we face a problem analogous to the integration of demographic noise in one dimension. The stochastic differential equation $\dot x = ax - x^2 + \sqrt{x} \xi(t)$ has an absorbing barrier at $x=0$, so the system stops if zero is reached. When the equation is integrated for small or negative $a$, the stationary state is $x^* \ll 1$. At this point, a large discrete fluctuation might push the system from $x(t)>0$ to $x_{t+\Delta t} < 0$ in a timestep, so the square root multiplicative noise cannot be further evaluated. Naively, one might think that once $x$ crosses to negative values, the absorbing barrier $x=0$ has been reached and the dynamics ends. However, if we repeat the same experiment, but with a smaller $\Delta t$, the strength of the stochastic fluctuation is reduced, thus reducing the probability of crossing to negative values and allowing the system to survive. Since any small amount of $x$ can trigger an "avalanche" of activity\cite{henkel_non-equilibrium_2008}, determining exactly when the absorbing state is reached is essential. It is well-known that just capping the value of $x$ at zero biases the result slightly \cite{dornic_integration_2005}.

Our problem is of the same nature, but instead of having negative values inside a square root, we produce a matrix that is not positive-definite. In the case of demographic noise, an exact integration algorithm is available \cite{dornic_integration_2005, weissmann_simulation_2018}, which prevents this problem. In our case, there is no choice but to "cap" the matrix to be positive definite.

The idea goes as follows: first of all, we obtain the normalized correlation matrix $\hat C$ from the covariance matrix, $\hat M$. This can be done defining $D_{ij} = \delta_{ij} \sqrt{M_{ij}}$, with $\delta_ij$ the Kronecker delta, and computing $\hat C = \hat D ^ {-1} \hat M \hat D ^ {-1}$.  One of the most efficient ways to determine if a matrix is positive-definite is to perform the Cholesky decomposition itself.  Thus, the matrix decomposition is attempted. If it succeeds, the matrix is positive definite, so we can recover the covariance matrix again by the inverse transform $\hat M = \hat D \hat C \hat D$. If not, we find the nearest correlation matrix $\hat {\tilde C}$ to $\hat C$. Then, we compute the new covariance $\hat {\tilde M} = \hat D \hat {\tilde C} \hat D$ and use it to continue.

Observe that --as it happened with the demographic noise-- the Cholesky decomposition tends to fail only when the system starts getting close to the absorbing barrier of full synchronization $R_k = 1$ and the timestep is large. The procedure above guarantees, however, that systems close to full synchronization with a reasonable step can be successfully integrated for a long time. As a downside, such a procedure introduces a small bias in the statistics of the fluctuations which is difficult to estimate. For the Kuramoto model, luckily, criticality happens at $R_k = 0$, so the system can be integrated at the transition without problems.

For the integration of systems with finite size, another problem arises: the factor $1/R_k$. The amplitude of the $k$-th mode vanishes in the limit of $k\to+\infty$. In practice, modes such as $k=10$ can have extremely small values of $R_k$ that will make the integration algorithm unstable. The problem is worsened by the computation of correlation matrices with very large entries. If one insists on evaluating the amplitude equation, the best is to assume that terms of order $1/N$ vanish, which removes the singularity, but keeps the fluctuations. A very effective and simple option, which is used in this work, is to keep the finite-size drift only for the first harmonic, neglecting it on the others. This constrains the Kuramoto order parameter to go to a fixed point above zero, thus driving the rest of the harmonics correctly.

Other small details need to be addressed. As with demographic noise, $R_k$ can cross to (unphysical) negative values, due to fluctuations, so one must enforce all $R_k$ to be positive. Since there is no absorbing state at $R_k=0$, this does not pose a big problem as in the demographic noise case, but it is better to enforce it for numerical stability. 

The divergence problem can be avoided by considering the equations for $x_k$ and $y_k$, the real and imaginary parts of the Kuramoto-Daido parameters, prior to the change of variables. No divergent term is present here. The price to pay is that then $2k$ equations have to be simulated, implying also larger covariance matrices. Since a Cholesky factorization (and, potentially, finding the nearest correlation matrix) is involved at each timestep, the numerical method is slower. 

A final highlight on the simulations: observe that in a closure with $k^*$ harmonics, $2k^*$ order parameters are needed to determine the covariance matrix. Setting all $R_{k>k^*} = 0$ is thus an important approximation. However, one can first solve the deterministic differential equations to find the numerically stable fixed point for $2k^*$ harmonics. Then, these values are used to determine the covariance matrix once and apply the Cholesky decomposition to compute  $\hat g$. This is guaranteed to work since no stochastic fluctuation was involved so far (see also Appendix F). Taking the stationary state as the initial condition, one simulates the equations with the fixed $\hat g$ previously obtained. This is a \emph{linear noise approximation}, which assumes that the fluctuations around the stationary state will be constant. This method is very accurate and way faster, but cannot be applied close to critical transitions --where no stationary state is well-defined-- and it was not used in the present paper. Nevertheless, it is present in the code repository. 

\subsection*{Implementation design}

All code was written in Julia and will be publicly available in a GitHub repository soon. The Kuramoto model was implemented by coupling each single oscillator to the mean field with an Euler-Milstein method and a timestep $\Delta t=0.01$. To measure the Kuramoto-Daido parameters, one can exploit the fact that  $\cos(k\phi)$ and $\sin(k\phi)$ can be written as a function of $\cos((k-1)\phi)$ and $\sin((k-1)\phi)$. Hence, it suffices to compute the first sine and cosine and obtain the Kuramoto-Daido parameters until the desired $k$ recursively. Since sum and multiplications are way faster than computing trigonometric functions, this turns out to be a noticeable optimization.

For the mesoscopic system, Julia allows us to integrate the differential equations efficiently while performing all matrix operations --such as Cholesky decomposition-- comfortably.  The nearest correlation matrix was implemented via a Newton method \cite{qi_quadratically_2006} readily implemented in \verb|NearestCorrelationMatrix.jl|. 
There is a different script for each one of the main cases discussed in the paper, to make the code easy to maintain and to facilitate replicability.

\begin{table}
\label{tab:parameters}
\begin{tabular}{c|c|c}
Parameter & Description & Value\tabularnewline
\hline 
$\Delta t$ & Timestep of Euler or Euler-Milstein integrator. & $0.01$\tabularnewline
$\omega$ & Frequency of the oscillators & $0.1$\tabularnewline
$\sigma^{2}$ & Noise intensity, location of critical point & $0.1$\tabularnewline
$k^{*}$ & Number of harmonics integrated & $30$\tabularnewline
$T_{r}$ & Relaxation time prior to start measuring & $4000$\tabularnewline
$T_{f}$ & Time for which the simulation is run. & $3\cdot10^{4}$\tabularnewline
$T_{m}$ & Time interval at which measures are taken & $1$\tabularnewline
$M$ & Number of simulations to average over & $100$\tabularnewline
\end{tabular}
\caption{\textbf{Parameters used by default in simulations.} These are the values typically used across the paper. When different values are used, it is stated in the corresponding figure caption.} 
\end{table}
In the case of the Kuramoto model, we start with random initial conditions for the oscillators. For the mesoscopic theory, the corresponding initial condition is $Z_k = 0$ for all the harmonics. The system is first integrated for a relaxation time $T_r$, and then measures are taken for a time $T_f$. Measurements are not taken at each timestep, but separated by a time $T_m$. Total integration time is $T_r + T_f$.  When computing a phase diagram, the coupling $J$ is gradually increased, and the state of the last simulation is used as initial condition to minimize relaxation time (which is fixed to $0.25T_f$). Each simulation is repeated $M$ times, and reported results are the average over $M$. For the histograms of order parameters, the values of all simulations are combined. 

The parameters used by default in all simulations are given in table \ref{tab:parameters}. These apply to the whole paper unless otherwise stated.

\section{Some properties of the amplitude covariance matrix}

For our simulation method, I have used that our covariance matrix is positive semidefinite when evaluated at the stationary solution. Also, in the main text, I used that inside the asynchronous phase the correlation matrix becomes diagonal, with additive noise. In this section, I review in detail these useful properties by analysing the correlation matrix of the amplitudes $R_k$.

The covariance matrix has the shape

\begin{equation}
    D_{kk'} = \frac{\sigma^2}{2N} k k' (R_{k-k'} - R_{k+k'})
\end{equation}

where the amplitude $R_{-k} = R_{k}$ due to the properties of the Kuramoto-Daido parameters. Additionaly, $R_0 = 1$.

First of all, it is simple to justify that in the incoherent regime the noise becomes additive. In such a regime, all the order parameters $R_k = 0$, and hence in the correlation matrix only the terms with $k=k'$ survive. The resulting matrix is diagonal and has the shape $D_{kk'} = \frac{\sigma^2 k^2}{2N}$, meaning that all the fluctuation variables become independent. 

Another interesting property, not so trivial to see, it whether $\hat D$ is positive definite or not. If the matrix is not positive definite when evaluated in the stationary state of the system it would indicate an inconsistency. I have checked this numerically, but here I offer an analytical argument. Formally speaking, I should substitute $R_k$ by the exact solution found in the main text with Bessel functions. However, for the sake of finding closed analytical solutions, I will refer to the approximate Ott-Antonsen ansatz.

There are several strategies to see if our matrix is positive definite.  One is to directly perform the Cholesky decomposition, as if it succeeds, the matrix is positive definite. Surprisingly, for the Ott-Antonsen form of the covariance matrix, 

\begin{equation}
    D_{kk'} = a k k' (R^{k-k'} - R^{k+k'}),
\end{equation}

the Cholesky decomposition can be done analytically. The first term is given by $g_{11} = \sqrt{D_11} = \sqrt{a} \sqrt{1-R^2}$. Then, the first column $g_{k1} = D_{k1} / g_{11} = \sqrt{a} kR^{k-1}\sqrt{1-R^2}$. From these, $g_{22} = \sqrt{D_{22} - g_{21} ^2} = 2\sqrt{a}\sqrt{1-R^2}$. The pattern starts to be apparent. Computing by hand a few terms immediately reveals that 

\begin{align}
    g_{kk'} = k\sqrt{a}R^{k-k'}\sqrt{1-R^2} \quad & k \leq k', \nonumber \\
    g_{kk'} = 0 \quad &k > k'.
\end{align}

which can be checked by multiplying the matrix $\hat g$ by its transposed. In this way, for the Ott-Antonsen ansatz it is possible to integrate the whole system immediately, without fearing for numerical errors. Since we were successful in computing the Cholesky decomposition, the Ott-Antonsen solution always leads to a positive-definite solution. It would remain to repeat this computation with the Bessel functions of the full recurrence, but numerical evidence seems to confirm my calculations in the general case.